%% file: main.tex
\documentclass[twocolumn,superscriptaddress,showpacs,prd,aps,amsmath,amssymb,nofootinbib]{revtex4-1}
\usepackage{multirow}
\usepackage{graphicx}
\usepackage{longtable}

\input kigou.tex

\usepackage[normalem]{ulem}
\usepackage{epstopdf}
\usepackage{color}
\usepackage{soul}
\usepackage{ulem}
\usepackage{bm}
\usepackage{xspace}

\usepackage[]{amsmath}
\usepackage[]{amsfonts}
\usepackage[]{amssymb}
\usepackage[]{mathrsfs}
\usepackage[]{mathtools}
\usepackage{times}
\usepackage{hyperref}

\hypersetup{
  colorlinks=true,        
  linkcolor=blue,         
  citecolor=cyan,         
}

\newcommand{\apjl}{Astrophys. J. Lett.}
\newcommand{\mnras}{Mon. Not. R. Astron. Soc.}
\newcommand{\nphysa}{Nucl. Phys. A}
\newcommand{\aap}{Astron. and Astrophys.}

\def\QEQ{{%
			\setbox0\hbox{$I$}%
			\rlap{\hbox to \wd0{\hss--\hss}}\box0
		}}

\everymath{\displaystyle}
\begin{document}
\title{Constraints of the maximum mass of quark stars based on post-merger evolutions}
\author{Yurui Zhou}
\affiliation{Department of Astronomy, School of Physics, Huazhong University of Science and Technology, Wuhan, 430074, P. R. China}

\author{Chen Zhang}
\affiliation{The HKUST Jockey Club Institute for Advanced Study,
The Hong Kong University of Science and Technology, Hong Kong, P. R. China}

\author{Junjie Zhao}
\affiliation{Henan Academy of Sciences, Zhengzhou 450046, Henan, China}
\affiliation{Department of Astronomy, Beijing Normal University, Beijing 100875, China}

\author{Kenta Kiuchi}
\affiliation{Max Planck Institute for Gravitational Physics (Albert Einstein Institute), Potsdam, D-14476, Germany}
\affiliation{Center for Gravitational Physics and Quantum Information, Yukawa Institute for Theoretical Physics, Kyoto University, Kyoto, 606-8502, Japan}

\author{Sho Fujibayashi}
\affiliation{Frontier Research Institute for Interdisciplinary Sciences, Tohoku University, Sendai 980-8578, Japan}
\affiliation{Astronomical Institute, Graduate School of Science, Tohoku University, Sendai 980-8578, Japan}
\affiliation{Max Planck Institute for Gravitational Physics (Albert Einstein Institute), Potsdam, D-14476, Germany}

\author{Enping Zhou}
\email{ezhou@hust.edu.cn}
\affiliation{Department of Astronomy, School of Physics, Huazhong University of Science and Technology, Wuhan, 430074, P. R. China}

\date{\today}
\begin{abstract}
We semi-analytically investigate the post-merger evolution of the binary quark star merger. The effective-one-body method is employed to estimate the energy and angular momentum dissipation due to gravitational waves in the inspiral phase. Three major mechanisms of energy and angular momentum dissipation are considered in the post-merger phase: mass outflows, neutrinos, and gravitational waves. The proportion of each mechanism could be determined by baryon number, energy and angular momentum conservation laws as well as the equilibrium model for rotating quark stars. Applying this analysis to the GW170817 event suggests two important conclusions: 1) a remnant quark star whose mass is smaller than the maximum mass of a uniformly rotating quark star can collapse before its rotational energy is dissipated via electromagnetic radiation (i.e., $\sim 100\,\mathrm{s}$) as the angular momentum left in the remnant quark star might not be large enough to sustain the additional self-gravity of the supramassive quark star due to the angular momentum dissipation of mass outflows, neutrinos and gravitational waves; 2) considering a general quark star equation of state model, a constraint on the maximum mass of cold and non-rotating quark stars is found as $M_{\mathrm{TOV}}\lesssim2.35^{+0.07}_{-0.17}\,M_{\odot}$, assuming a delayed collapse occurred before a large fraction of the total rotational energy ($ \gtrsim 10^{53}\,$erg) of the merger remnant was deposited into the merger environment for the GW170817 event. These constraints could be improved with future merger events, once there are more evidences on its post-merger evolution channel or information on the amount of post-merger gravitational wave and neutrino emissions inferred from the multi-messenger observations.
\end{abstract}

\maketitle

\section{INTRODUCTION}
\label{sec:introduction}

The first direct detection of gravitational waves from an inspiraling binary system of two neutron stars (GW170817)~\citep{Abbott2017} followed by a wide variety of the observations of electromagnetic counterparts (GRB 170817A, AT 2017gfo)~\citep{Abbott2017b} have enriched our comprehension of the equation of states (EOS) of neutron stars (NSs). Gravitational wave observations for the late inspiral phase of binary neutron stars (BNS) can be used to constrain the tidal deformability of NSs ~\citep{Abbott2018,Abbott2019,Annala2018,Soumi2018}. 
Moreover, the electromagnetic counterparts serve as indicators of the fate of the merger remnant, which can also be used to constrain the EOS of merging NSs~\citep{Shibata2017,Bauswein2017,Margalit2017,Ruiz2018,Rezzolla2018,Shibata2019}.

The property of the blue kilonova component of AT 2017gfo disfavors the case of prompt collapse, i.e., a black hole (BH) is formed on a dynamical timescale following the merger \citep{Kasen2017,Margalit2017}, leading to a constraint on the radius of NSs \citep{Bauswein2017}.
The case of a stable NS remnant is also disfavored, as there is no observational evidence of significant energy injection from the remnant NS. Therefore, a delayed collapse to BH scenario is believed to happen following the merger of GW170817 event, and hence an upper limit of the maximum mass of non-rotating NSs is obtained $M_{\mathrm{TOV}}\lesssim 2.2\,M_{\odot}$ \citep{Margalit2017, Rezzolla2018}. A similar constraint can be obtained from the observation of GRB 170817A \citep{Ruiz2018}.
However, it is pointed out that at the onset of collapse, the merger remnant NS is not necessarily at the mass-shedding limit \citep{Shibata2019}, which is an assumption that many previous studies applied when deriving the constraints on $M_{\mathrm{TOV}}$. 
Employing both energy and angular momentum conservation laws as well as results of numerical-relativity (NR) simulations, it has been suggested that $M_{\mathrm{TOV}}$ could be as large as $\sim 2.3 \, M_{\odot}$ \citep{Shibata2019}.

Nevertheless, the EOS of NSs is currently a subject of ongoing debate, primarily due to the complex non-perturbative QCD problems associated with it. In addition to conventional NS models, recent studies have suggested the possibility of the existence of deconfined quark matter in the high-density core of massive NSs \citep{Annala2020}. Numerous efforts have been made to detect evidence of a transition from conventional hadronic matter to quark matter, a strong-interaction phase transition, within NSs through astrophysical observations \citep{Bauswein2019,Most2019}.  Alternatively, the existence of quark stars (QSs), either strange quark stars (SQSs) composed entirely of deconfined $ u,d,s $ quarks or up-down quark stars ($ud$QSs) composed entirely of $u, d$ quarks, has also been proposed as a potential explanation for the nature of compact stars~\citep{Alock1986,Drago2016,Drago2016b,Bhattacharyya2017,Zhang:2019mqb,Wang:2019jze,Ren:2020tll},
based on the conjecture that either strange quark matter (SQM) comprising $ u,d,s $ quarks~\citep{Bodmer1971,Witten1984} or up-down quark matter ($ud$QM) comprising $ u,d$ quarks~\citep{Holdom:2017gdc} could represent the true ground state of the bulk strong matter. It has been shown that interquark effects, such as color superconductivity and QCD corrections, could significantly increase the $M_{\rm TOV}$ of QSs to explain the mass-gap object ($M\sim 2.6 M_{\odot}$) detected in GW190814~\citep{Zhang2021,Roupas:2020nua,Horvath:2020cjz,Albino:2021zml,Miao:2021nuq,Restrepo:2022wqn,Oikonomou:2023otn,Cao:2020zxi}, which cannot be naturally accommodated by NS considering NSs' $M_{\rm TOV}\lesssim2.3 M_{\odot}$ as constrained by the post-merger NS evolution in GW170817 event~\citep{Shibata2019}. However, the post-merger evolutions of quark stars, if assuming GW170817 is a merger of binary quark stars (BQS), could also similarly set a tight upper bound on QSs' $M_{\rm TOV}$, and this essential point has been commonly overlooked in the literature. 

Previous studies have shown that most of the matter ejected
during and after the merger of BQS could evaporate
into nucleons and therefore BQS merger can possibly generate a kilonova-like signal \citep{Paulucci2017,Bucciantini2019}. Based on recent astrophysical observations, constraints have been established on the EOS for QSs \citep{Zhou2018,Zhang2021}. Efforts have also been undertaken to 
model the dynamics \citep{Bauswein2010,Zhou2022} and electromagnetic counterparts \citep{Bauswein2009,Li2016,Paulucci2017,Lai2018,Bucciantini2019} associated with BQS mergers.

It is known that uniform rotation significantly increases the maximum mass of QSs in comparison to NSs (i.e., the ratio of the maximum mass of uniformly rotating NSs to the maximum stable mass of non-rotating NSs  $M_{\mathrm{max,urot}}/M_{\mathrm{TOV}}$ is typically 1.2 for NSs, but 1.4 for QSs). This difference may make a distinction in the scenario of the post-merger outcome. Specifically, due to the discovery of pulsar more massive than $2\,M_{\odot}$ \citep{Fonseca2021, Cromartie2020}, the mass of a QS at the mass-shedding limit of rigid rotation is at least $2.8\,M_{\odot}$, which exceeds the total mass of the system in GW170817 event ($M=2.74^{+0.04}_{-0.01} \, M_{\odot}$ with the $90\%$ credible level \citep{Abbott2017}). Therefore, for the BQS case, a long-lived QS should be formed after the merger of the GW170817 event.
However, as pointed out in \citep{Shibata2019},
for the BNS case at the onset of collapse to a BH the merger remnant may not contain enough angular momentum to reach the mass-shedding limit, and hence the BQS merger remnant for the GW170817 event is not necessarily a long-lived QS if the situation is similar for the BQS case. Therefore, analysis similar to \citep{Shibata2019} needs to be done for the BQS case to understand the post-merger outcome of the BQS merger.

In this paper, we perform a similar analysis to \citep{Shibata2019} for the BQS scenario to investigate the post-merger evolution channels of the BQS mergers, specifically, we focus on the case of prompt collapse and delayed collapse. In comparison to \citep{Shibata2019}, the effective-one-body (EOB) approach \citep{Buonanno1999,Buonanno2000}, which enables accurate computation of gravitational waveforms throughout the inspiral phase, is employed in our study to estimate the energy and angular momentum dissipation during the inspiral stage, instead of applying the empirical relation between the dissipated energy and angular momentum obtained through NR simulations. 
In addition, a recent study \citep{Zhang2021} has derived an EOS for interacting quark matter (IQM) unifying all cases by a simple reparametrization and rescaling, through which the number of degrees of freedom in the QS EOS
can be maximally reduced. With this unified EOS, we systematically investigate the equilibrium models of rotating QSs and analyze the fate of the merger remnant to establish the constraint on IQM parameter space from observation.

This paper is organized as follows. In Sec.~\ref{sec:prompt}, we analyze the prompt collapse case for the BQS merger. In Sec.~\ref{sec:delayed}, we investigate the delayed collapse scenario. The constraint on the maximum mass of QSs from the GW170817 event is established in Sec.~\ref{sec:170817}. Finally, we summarize our work and present our conclusions in Sec.~\ref{sec:conclusion}. Throughout this paper, we employ the unis of $c = G = M_{\odot}=1 $ where $c, \,G $ and $ M_{\odot}$  are the speed of light, the gravitational constant and the solar mass, respectively.

\section{PROMPT COLLAPSE}
\label{sec:prompt}

The outcome of a BNS merger depends on the total mass of the system. A direct gravitational collapse to a BH takes place if the total mass is beyond a certain mass, which is called the threshold mass for prompt collapse ($M_{\mathrm{thres}}$) \citep{Shibata2005,Shibata2005b,Shibata2006}.  Therefore, $M_{\mathrm{thres}}$ is an important parameter to characterize the fate of the merger remnant. Many attempts have been made to determine $M_{\mathrm{thres}}$ for the scenario of BNS mergers \citep{Bauswein2013,Bauswein2017b,Bauswein2020,Bauswein2021,koppel2019,Kiuchi:2010ze}, while only a limited number of studies for the BQS case  \citep{Zhou2022} due to the difficulties in hydrodynamical simulations of BQS mergers arising from the strong density discontinuity at the stellar surface. Hence, to investigate $M_{\mathrm{thres}}$ of a large number of different QS EOSs, a method with relatively small computational demands is needed. It has been suggested in \citep{Bauswein2017b} that whether
or not prompt collapse occurs depending mostly on the relation between the mass of the merger remnant and the maximum stable mass ($M_{\mathrm{stab}}$) given the angular momentum contained in the remnant at the time of the merger ($J_{\mathrm{merger}}$). $M_{\mathrm{stab}}$ can be approximately derived from differentially rotating equilibrium models \cite{Bauswein2017b,Lukas2018,Zhou2019}. Therefore, through this semi-analytical method, one can derive $M_{\mathrm{thres}}$ without time-consuming and sophisticated hydrodynamical simulations. 

In this section, we apply the method proposed in \citep{Bauswein2017b} to determine $M_{\mathrm{thres}}$ for the case of BQS mergers. In addition, the empirical relation between $M$ and $J_{\mathrm{merger}}$ constructed through NR simulations is used in \citep{Bauswein2017b} to estimate the angular momentum of the merger remnant. Compared to the empirical relation, the EOB method employed in this paper permits a more accurate calculation of the angular momentum left in the merger remnant, and hence a more precise $M_{\mathrm{thres}}$ can be derived, which can be used to constrain QS EOSs only relying on the measurement of the total mass of the binary and the determination of whether a prompt collapse takes place or not in the event.

\subsection{EOS models}
\label{sec:eos}
In~\citep{Bauswein2013,Bauswein2017b,Bauswein2020,Bauswein2021}, various NS EOS models are constructed to derive the EOS-insensitive relations between $M_{\mathrm{thres}}$ and stellar properties. For QSs, it is known that in the simplest MIT bag model (i.e., with zero strange quark mass and neglecting interaction between quarks), stellar properties of both rotating and non-rotating QSs scale with the bag constant which is also the only parameter describing the EOS model. Therefore, the simplest MIT bag model is not appropriate for exploring the parameter space and deriving the EOS-insensitive relations for $M_{\mathrm{thres}}$ of the BQS merger. In this paper,  we employ the unified QS EOS including the perturbative QCD corrections~\citep{Fraga2001,Alford2005} and color superconductivity~\citep{Rajagopal2001,Lugones2002,Alford1999} derived in \citep{Zhang2021}, where with a simple reparametrization and rescaling, the EOS can be expressed as (see \citep{Zhang2021} for details):

\begin{equation}
    p=\frac{1}{3}(\rho-4B_{\mathrm{eff}})+\frac{4\lambda^2}{9\pi^2}\left( -1 + \mathrm{sgn}(\lambda) \sqrt{1+3\pi^2 \frac{\rho -B_{\mathrm{eff}}}{\lambda^2}}\right) \label{14}
\end{equation}
where $\mathrm{sgn}(\lambda)$ represents the sign of $\lambda$, $p$ denotes the pressure, $\rho$ corresponds to the total energy density and $B_{\mathrm{eff}}$ is the effective bag constant as a phenomenological representation
of non-perturbative QCD vacuum effects. $\lambda$ is defined
as:
\begin{equation}
    \lambda=\frac{\xi_{2a}\Delta^2-\xi_{2b}m_s^2}{\sqrt{\xi_4 a_4}},
\end{equation}
where $m_s$ represents the finite mass of the strange quark, $\Delta$ is the pairing energy gap arising from color superconductivity and $a_4$ parametrizes the perturbative QCD corrections. $\xi_{2a}$, $\xi_{2b}$ and $\xi_4$ are the constant coefficients and can be expressed as:

\begin{equation} (\xi_4, \xi_{2a},\xi_{2b})=
\begin{cases}
(((\tfrac{1}{3})^{\frac{4}{3}}+\tfrac{2}{3})^{\frac{4}{3}})^{-3},1,0) & \mathrm{2SC\ phase} \\
(3,1,3/4) & \mathrm{2SC+s\ phase}\\
(3,3,3/4) & \mathrm{CFL\ phase}
\end{cases}, 
\label{phases}
\end{equation}
for various phases of quark matter. Here the 2SC and 2SC+s phases both represent the two-flavor color superconductivity phase where $u$ quarks pair with $d$ quarks, while only the latter has (unpaired) strange quarks in composition. CFL phase denotes the color-flavor locking phase, where $u, d, s$ quarks pair with each other antisymmetrically.

By introducing the following dimensionless rescaling:

\begin{equation}
    \Bar{\rho}=\frac{\rho}{4B_{\mathrm{eff}}},\  \Bar{p}=\frac{p}{4B_{\mathrm{eff}}}, \ \Bar{\lambda}=\frac{\lambda^2}{4B_{\mathrm{eff}}},
    \label{15}
\end{equation}
Eq.~(\ref{14}) can be expressed as
\begin{equation}
    \bar{p}=\frac{1}{3}(\bar{\rho}-1)+\frac{4\Bar{\lambda}}{9\pi^2}\left( -1 + \mathrm{sgn}(\lambda) \sqrt{1+\frac{3\pi^2}{\Bar{\lambda}} (\Bar{\rho}-\frac{1}{4}})\right). \label{16}
\end{equation}
In Eq.~(\ref{16}), $\mathrm{sgn}(\lambda)\,\bar{\lambda}$ is the only parameter governing the EOS models, and thus we can obtain different families of QS EOS by selecting different $\mathrm{sgn}(\lambda)\,\bar{\lambda}$. In addition, for every EOS family with a certain value of $\mathrm{sgn}(\lambda)\,\bar{\lambda}$, the solution scales with $B_{\mathrm{eff}}$ and the properties of a QS such as the gravitational mass and the radius can be obtained given a specific $B_{\mathrm{eff}}$, just like the case of MIT bag model. In this paper, to construct a series of representative EOSs, we explore a broad range of values for $\bar{\lambda}$, ranging from $0$ (indicating non-interacting QSs) to $10$ (representing a high level of strong interaction strength). In this paper, we only concentrate on the case of the CFL phase, with the employed benchmark EOSs listed in Table  \ref{tab:eos}. Our analyses and results should also apply to the 2SC+s phase with the same values of ($\mathrm{sgn}(\lambda)\,\bar{\lambda}$, $B_{\mathrm{eff}}$, $a_4$), since the only difference between these two phases is the value of $\xi_{2a}$, which does not influence the relation among pressure, energy density and number density when $\mathrm{sgn}(\lambda)\,\bar{\lambda}$, $B_{\mathrm{eff}}$ and $a_4$ are fixed~\citep{Zhang2021}.
\begin{table*}[t]

    \centering
	\begin{tabular}{c@{\hskip 10pt}c@{\hskip 4pt}c@{\hskip 10pt}c@{\hskip 8pt}c@{\hskip 6pt}c@{\hskip 4pt}c@{\hskip 4pt}c@{\hskip 10pt}c@{\hskip 10pt}c@{\hskip 15pt}c@{\hskip 15pt}c@{\hskip 6pt}c}
		\hline
		\hline
           Model & $\Delta$ (MeV) & $m_s $ (MeV)& $a_4$ & $B_{\mathrm{eff}}^{(1/4)}$ (MeV) & $\mathrm{sgn}(\lambda)\,\bar{\lambda}$ & $M_{\mathrm{TOV}}$ ($M_{\odot}$) & $R_{1.4}$ (km)&$R_{\mathrm{TOV}}$ (km)& $\Lambda_{1.4}$ & $f_0$ & $f_{\mathrm{TOV}}$ & $f_{\mathrm{peak}}$ (kHz) \\
        \hline
            EOS-1 & 137.78 & 100.00 & 0.50 & 168.96 & 0.500 & 2.013 & 10.13 &10.27& 286 & 1.162 & 1.238 & 3.33\\ 
            EOS-2 & 0.00  & 100.00 & 0.61 & 137.16 & -0.022 & 2.080 & 11.41 & 11.47& 619 & 1.132 & 1.201 & 2.70\\
            EOS-3 & 0.00  & 0.00   & 0.57 & 141.70 & 0.000      & 2.100 & 11.33& 11.45 & 598 & 1.132 & 1.202 & 2.73\\
            EOS-4 & 0.00  & 100.00 & 0.61 & 135.14 & -0.023 & 2.137 & 11.66 &11.83 & 718 & 1.144 & 1.217 & 2.59\\
            EOS-5 & 160.85 & 100.00 & 0.50 & 169.19 & 1.000 & 2.200 & 10.52 &11.13& 382 & 1.251 & 1.351 & 3.11 \\
            EOS-6 & 151.41 & 100.00 & 0.40 & 167.23 & 1.000 & 2.254 & 10.70 &11.35& 429 & 1.194 & 1.293 & 3.01 \\
            EOS-7 & 100.00& 100.00 & 0.61 & 148.51 & 0.142 & 2.284 & 11.56 &12.04 & 703 & 1.233 & 1.327 & 2.61\\
            EOS-8 & 228.51 & 100.00 & 0.13 & 194.33 & 10.000 & 2.284 & 9.50 &10.15& 195 & 1.190 & 1.314 & 3.66\\
            EOS-9 & 97.83 & 100.00 & 0.50 & 147.13 & 0.160 & 2.350 & 11.78 & 12.35 & 796 & 1.190 & 1.284 & 2.52\\
            EOS-10 & 148.77 & 100.00 & 0.40 & 163.95 & 1.000 & 2.350 & 11.02 &11.70& 524 & 1.214 & 1.321 & 2.84\\

        \hline
		\hline
	\end{tabular}
	\caption{The EOS of interacting QSs used in this paper, and the important parameter of non-rotating QSs. $\Delta$ is the pairing energy gap arising from color superconductivity. $m_s$ is the mass of the strange quark, $a_4$ is the perturbative QCD correction parameter.  $B_{\mathrm{eff}}$ is the effective bag constant including a phenomenological representation of non-perturbative QCD vacuum effects. $\bar{\lambda}$ is the quantity that characterizes the strength of the related strong interaction defined in \citep{Zhang2021}. $M_{\mathrm{TOV}}$ and $R_{\mathrm{TOV}}$ are the mass and radius of maximum mass non-rotating QSs. $R_{1.4}$ and $\Lambda_{1.4}$ are the radius and the tidal deformability of 1.4 $M_{\odot}$ QSs. $f_0$ represents the ratio of the baryonic mass
    to the gravitational mass of  $1.37\,M_{\odot}$ QSs. $f_{\mathrm{TOV}}$ is the ratio of the baryonic mass to the gravitational mass for the maximum mass non-rotating QSs. $f_{\mathrm{peak}}$ is the peak frequency in the spectrum amplitude of gravitational waves emitted in the post-merger phase estimated by Eq. (5.4) of Ref. \citep{Kiuchi2020}. }
	\label{tab:eos}
    
\end{table*}

QSs with the EOSs listed in Table \ref{tab:eos} are capable of reproducing $2.0\, M_{\odot}$ QSs \citep{Fonseca2021, Cromartie2020}, and also satisfying the GW170817 constraint $\Lambda_{1.4} \lesssim 800$~\citep{Abbott2017,Soumi2018}, where $\Lambda_{1.4}$ is the tidal deformability of a $1.4\, M_{\odot}$ QS. 
It is worth noting that the relation between energy density and pressure is solely dependent on $(\mathrm{sgn}(\lambda)\,\bar{\lambda}, B_{\mathrm{eff}})$. Consequently, the values of $M_{\mathrm{TOV}}$ and $\Lambda_{1.4}$ are determined solely by $(\mathrm{sgn}(\lambda)\,\bar{\lambda}, B_{\mathrm{eff}})$ and
the parameters encoding details of nuclear physics (i.e., $\Delta,m_s$ and $a_4$) listed in Table \ref{tab:eos} can take any values as long as $(\mathrm{sgn}(\lambda)\,\bar{\lambda}, B_{\mathrm{eff}})$ remain unchanged.

\subsection{Effective-one-body model}

During the inspiral phase, the angular momentum dissipation of the binary system is primarily due to gravitational waves. Therefore, the angular momentum left in the merger remnant can be expressed as:

\begin{equation}
    J_{\mathrm{merger}}=J_0-J_{\mathrm{GW,i}},  \label{1}
\end{equation}
where $J_0$ represents the angular momentum of the binary at the beginning of the calculation of the gravitational waves and $J_{\mathrm{GW,i}}$ corresponds to the angular momentum radiated by gravitational waves in the inspiral phase. 

In this paper, we just focus on the case of quasi-circular, nonspinning coalescences. In this case, given the properties of the binary, $J_0$ can be obtained through the EOB dynamics for a specific initial orbit frequency or a specific initial separation \citep{Buonanno2000}. When the initial condition is determined, the gravitational waveform until the time of the merger can be generated through the EOB approach. Moreover, the radiated angular momentum can be calculated as:

\begin{gather}
    J_{\mathrm{GW,i}}=\frac{1}{16\pi} \sum\limits_{(l,m)}\int_{t_0}^{t_{\mathrm{merger}}}dt^{\prime}m\Im[r^2h_{l m}\left( t^{\prime} \right) \dot h_{l m}^* \left( t^{\prime} \right) ], \label{2}
\end{gather}
where $h_{l m}$ is the gravitational waveform generated via the EOB method, $r$ represents the radius where $h_{l m}$ is extracted, $t^{\prime}$ corresponds to the time from the start of the simulation ($t_0$) to the merger ($t_{\mathrm{merger}}$), $\Im$ is the imaginary part of a complex number and $(l,m)$ represents the different modes of gravitational waves. The $x$ and $y$ components of angular momentum loss are not included because they turn out to have a negligible effect on the computation of $J_{\mathrm{GW,i}}$ \citep{Damour2012,Bernuzzi2012}. As shown in \citep{Dietrich2017b,Kiuchi2020}, most of the energy radiated in a BNS merger is contained in the (2,2) mode of gravitational waves. Therefore, for simplicity, only the dominant (2,2) mode is considered in this paper.

In the present paper, we employ the state-of-the-art time-domain EOB waveform model, specifically \texttt{SEOBNRv4T} \citep{Hinderer2016, Steinhoff2016}, which is available in the LSC Algorithm Library \citep{lalsuite}. 
\texttt{SEOBNRv4T} incorporates dynamic tides and accounts for the spin-induced quadrupole moment effects. This model aligns with the latest BNS numerical simulations, extending up to the merger stage \citep{Dietrich2017}, and has been currently implemented for LIGO data analysis \cite{Abbott2019}. Similar to \texttt{SEOBNRv4T}, Other accurate gravitational waveform models like \texttt{TEOBResumS}~\citep{Bernuzzi2015b,Nager2018,Akcay2019}, \texttt{NRTidalv2}~\citep{Dietrich2019}, \texttt{NRTidalv3}~\citep{Abac2024} can also be used to obtain $J_{\mathrm{merger}}$. The comparison among these gravitational waveform models~\citep{Abac2024} shows that the deviation of waveform appears only a few orbits before the merger and the dissipation in these few orbits is minor compared to the dissipation in the whole inspiral phase, and thus the systematic difference among these models is insignificant for calculating $J_{\mathrm{merger}}$. In the rest of this paper, we only show results calculated through \texttt{SEOBNRv4T} model and compare them with the results obtained by NR simulations to estimate the error of our method.

Given an EOS, we first set the mass components ($M_1$, $M_2$) of the binary and set the initial orbit frequency, and then calculate the tidal deformability of both NSs ($\Lambda_1$, $\Lambda_2$). After that, we can use \texttt{SEOBNRv4T} model to generate the gravitational waveforms, and $J_{\mathrm{GW,i}}$ can be calculated through Eq. (\ref{2}). As mentioned above, given the initial orbit frequency, $J_0$ can also be derived through Eqs. (4.5) and (4.8) in \citep{Buonanno2000}. Finally, $J_{\mathrm{merger}}$ can be obtained through Eq. (\ref{1}).

To evaluate the accuracy of our method, we make a comparison among the angular momentum at the moment of merger obtained by the EOB method, NR simulations, and empirical relations. Here we employ the results of NR simulations in \citep{Bernuzzi2016} and the empirical relation Eq. (3.1) proposed in \citep{Shibata2019}. Table \ref{tab:merger} lists the value of $J_{\mathrm{merger}}$ of different binary parameters derived via different methods. It can be seen that the values of $J_{\mathrm{merger}}$ obtained through the EOB method and NR simulations differ by approximately 1$\%$, while the deviation between empirical relation and NR simulations is about 5$\%$. Therefore, through the EOB method, $J_{\mathrm{merger}}$ can be obtained accurately without relying on the time-consuming NR simulations.

\begin{table*}[t]

    \centering
	\begin{tabular}{c|c@{\hskip 40pt}c@{\hskip 40pt}c@{\hskip 40pt}c@{\hskip40pt}c}
		\hline
		\hline
            EOS, references &  $M_1$  & $M_2$  &    $J_{\mathrm{merger}}^{\mathrm{EOB}} $   & $J_{\mathrm{merger}}^{\mathrm{NR}}$  & $J_{\mathrm{merger}}^{\mathrm{empirical}}$ \\
        \hline
            \multirow {4}{*}{DD2, \citep{Hempel2010,Typel2010} } & 1.4 & 1.2  & 5.904  & 5.996  & 5.944 \\
                                    & 1.365& 1.25  & 5.981 & 6.066  & 6.154\\
                                    & 1.35 & 1.35  & 6.33 & 6.379 & 6.633 \\
                                    & 1.44 & 1.39 & 6.853  & 6.888 & 7.274 \\
        \hline
            \multirow {4}{*}{SFHo, \citep{Steiner2013} } & 1.4 & 1.2   & 5.699  & 5.726  & 5.807 \\
                                    & 1.365& 1.25  & 5.773  & 5.792  & 6.010\\
                                    & 1.35 & 1.35 & 6.101  & 6.109 & 6.478 \\
                                    & 1.44 & 1.39  & 6.595  & 6.615 & 7.104 \\

        \hline
		\hline
	\end{tabular}
	\caption{The values $J_{\mathrm{merger}}$ obtained through different methods. The units of mass and angular momentum are $M_{\odot}$ and $GM^2_{\odot}/c$, respectively. Here $M_1$ and $M_2$ denote the individual isolation masses,  $J_{\mathrm{merger}}^{\mathrm{EOB}} $,   $J_{\mathrm{merger}}^{\mathrm{NR}}$ and $J_{\mathrm{merger}}^{\mathrm{empirical}}$ are the angular momentum at the moment of merging calculated by the EOB method, NR simulations and the empirical relation respectively. }
	\label{tab:merger}
    
\end{table*}

\subsection{Threshold mass of prompt collapse}
Given the total mass and angular momentum of the merger remnant, the maximum stable mass needs to be calculated to determine the threshold mass for prompt collapse. Similar to the previous study \cite{Bauswein2017b,Lukas2018,Zhou2019}, we estimate $M_{\mathrm{stab}}$ for given angular momentum through the differentially rotating equilibrium models.

\begin{figure}
	\begin{center}
		\includegraphics[height=70mm]{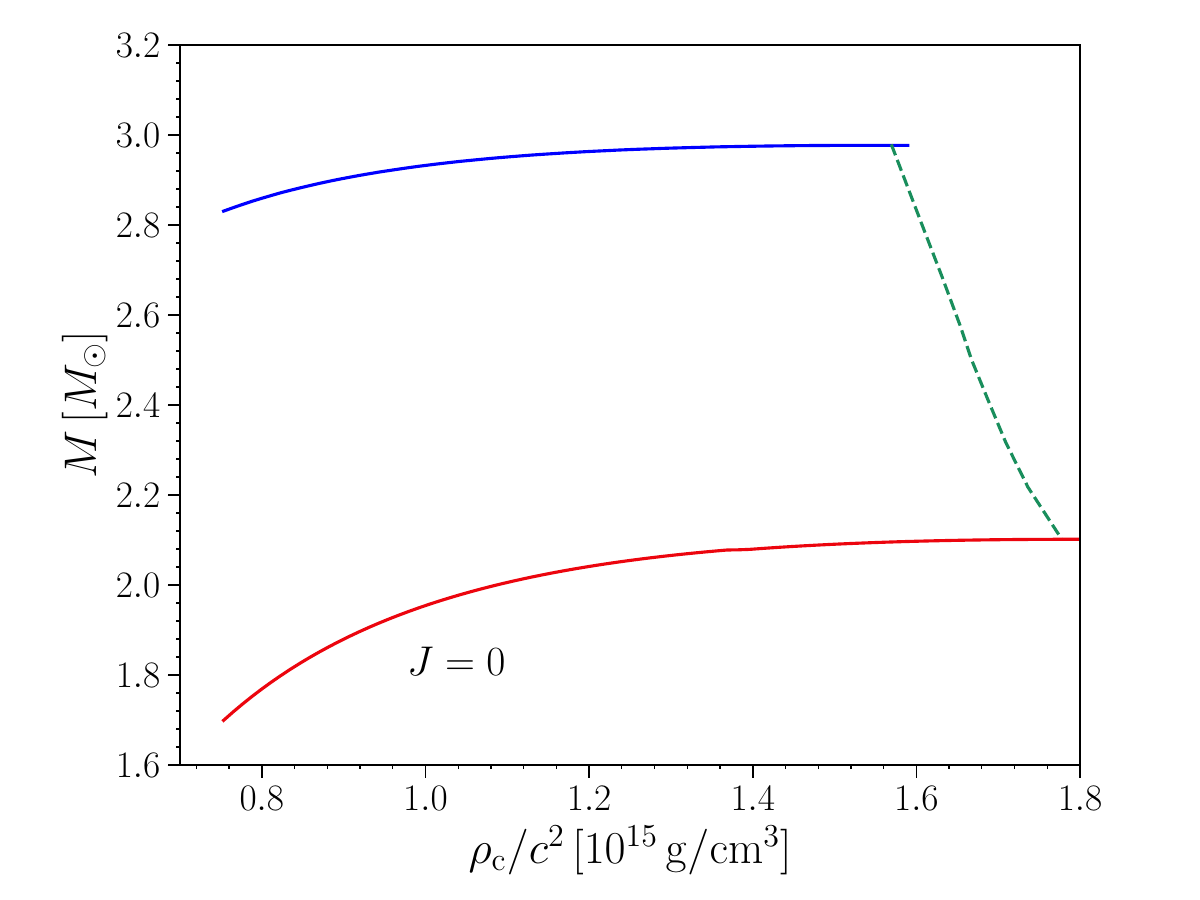}
	\end{center}
	\caption{ Gravitational mass versus central density diagram of uniformly rotating QSs with various fixed values of angular momentum $J$ for the EOS-3. The lower and upper solid curves show the sequences of non-rotating QSs (labelled by $J=0$) and uniformly rotating QSs at mass-shedding limits. The dashed curve represents the sequence of the turning points, where QSs are marginally stable to gravitational collapse. QSs located on the lower density side of this curve are stable, while those on the higher density side are unstable.}
	\label{fig:TP}
\end{figure}

The relation between $M_{\mathrm{stab}}$ and $J_{\mathrm{merger}}$ can be determined through the “turning point” theorem of Friedman, Ipser, and Sorkin \citep{Friedman1988}.
According to the research, for rigidly rotating stars, along a sequence with a constant angular momentum and varying mass and central energy density ($\rho_{\mathrm{c}}$), a secular instability occurs at the maximum mass, corresponding to the turning point of the $M-\rho_{\mathrm{c}}$ curve (see Figure \ref{fig:TP} for turning points). In addition, as shown in \citep{Lukas2018}, the turning point criterion can be seen as a first approximation for determining $M_{\mathrm{stab}}$ for differentially rotating stars.

The differentially rotating QSs have been investigated in \citep{Zhou2019}, where it has been indicated that the details of the rotational profile in the merger remnant only have little impact on the relation between the maximum stable mass and the angular momentum of the remnant. Therefore, for simplicity, we adopt the widely used differential rotation law known as the $j$-constant law \citep{Komatsu1989},

\begin{equation}
    j(\Omega)=A^2(\Omega_{\mathrm{c}}-\Omega),
\end{equation}
where $A$ is a constant parameter in the model, $j:=u^tu_{\phi}$ is the specific angular momentum, $\Omega:=u^\phi/u^t$ 
is the angular velocity as measured by an observer at infinity, and $\Omega_{\mathrm{c}}$ is the angular velocity at the center of the star. We choose $\hat{A}:=A/r_{\mathrm{e}}=$1.5, 2 and 3 respectively to estimate the differences of various rotation laws, where $r_{\mathrm{e}}$ is the equatorial radius of the star.

\begin{figure}
	\begin{center}
		\includegraphics[height=70mm]{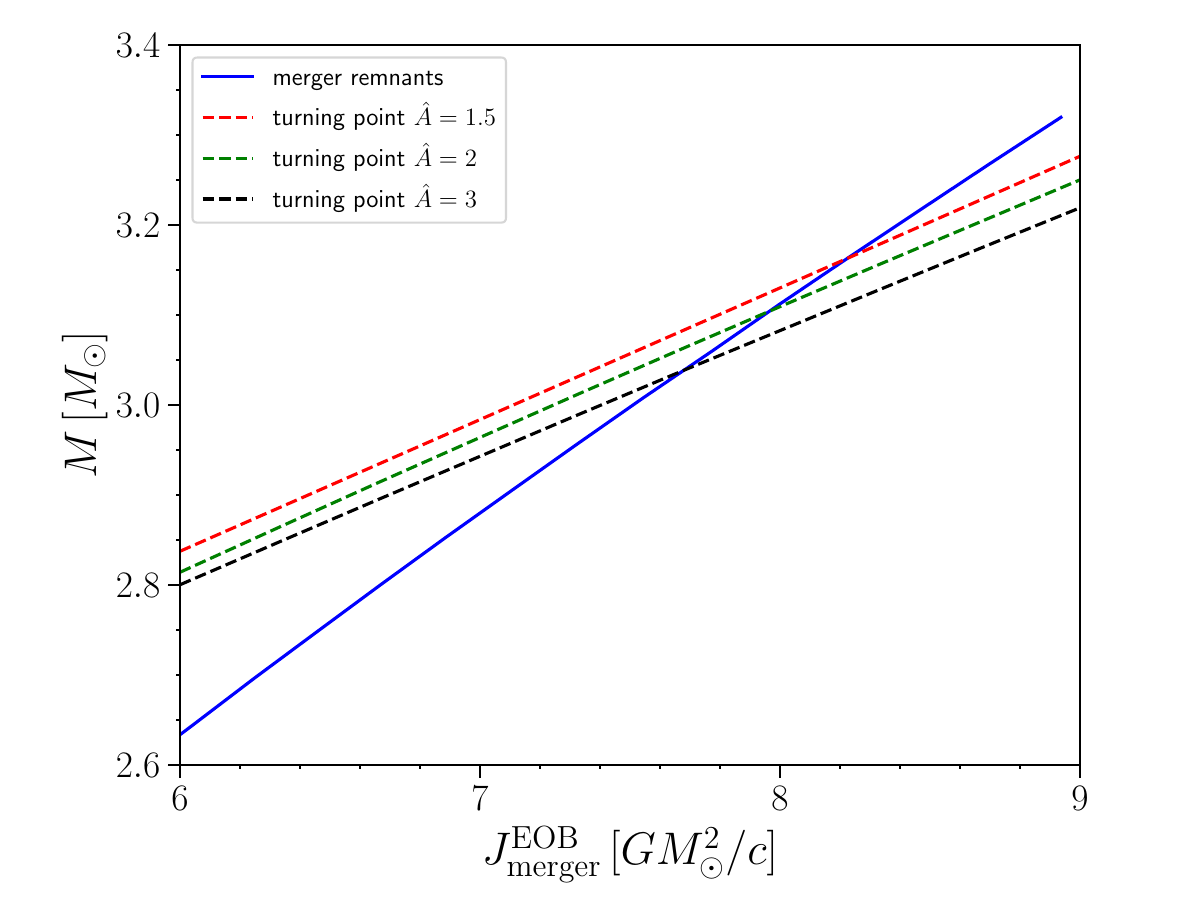}
   
	\end{center}
	\caption{Deriving the threshold mass to collapse by our equilibrium models. The dashed line corresponds to the relation between the maximum stable mass $M_{\mathrm{stab}}$ and the angular momentum $J$ of differentially rotating QSs with different rotation laws. The blue solid line represents the relation between the total binary ADM mass $M$  and the angular momentum of the remnant at the time of merger computed by the \texttt{SEOBNRv4T} model $J_{\mathrm{merger}}^{\mathrm{EOB}}$ for the equal-mass case. The intersection of these two lines determines the threshold mass to collapse calculated from our equilibrium models. } 
	\label{fig:thr}
\end{figure}

We first employ EOS-3 the same EOS as used in NR simulations \citep{Zhou2021,Zhou2022}. \texttt{RNS} code \citep{Stergioulas1995} is used to construct the equilibrium models of differentially rotating QSs. The relation between $M_{\mathrm{stab}}$ and $J$ of differentially rotating QSs with the $j$-constant law with $\hat{A}=$ 1.5, 2 and 3 calculated by \texttt{RNS} code and the relation between $M$ and $J_{\mathrm{merger}}$ computed by the EOB method are displayed in Figure \ref{fig:thr}. The point of intersection between the solid and dashed lines in Figure \ref{fig:thr} can be used to provide an estimation of the threshold mass of the prompt collapse in BQS mergers ($\approx 3.10^{+0.06}_{-0.06}\,M_{\odot}$). $M_{\mathrm{thres}}$ of this EOS has been determined through NR simulations \citep{Zhou2022} to range from $3.05\,M_{\odot}$ to $3.10 M_{\odot}$. The two results differ only by $\sim2\%$, which may be attributed to the assumptions of axisymmetry and stationarity, and the uncertainty in the differential rotation law used in constructing the equilibrium models. Compared to the results in \citep{Bauswein2017b}, it can be found that with the help of the EOB method, the deviation of $M_{\mathrm{thres}}$ derived by the equilibrium models decreases from $\sim5\%$ to $\sim2\%$. 

Moreover, as discussed in \citep{Zhang2021,Zhou2022}, both $M_{\mathrm{TOV}}$ and $M_{\mathrm{thres}}$ scale with the bag constant in the same way (see Eq. (4) in \citep{Zhou2022}). The quantities used in this approach ($M_{\mathrm{stab}}$ and $J$ for rotating QSs) can also scale with the bag constant \citep{Bhattacharyya2016}. Therefore, the consistency between the results obtained from this approach and NR simulations is not coincidental and the validation of this approach is independent of the choice of the bag constant. Consequently, this semi-analytical method allows the study of a large quantity of different QS EOSs.

\begin{figure}
	\begin{center}
		\includegraphics[height=70mm]{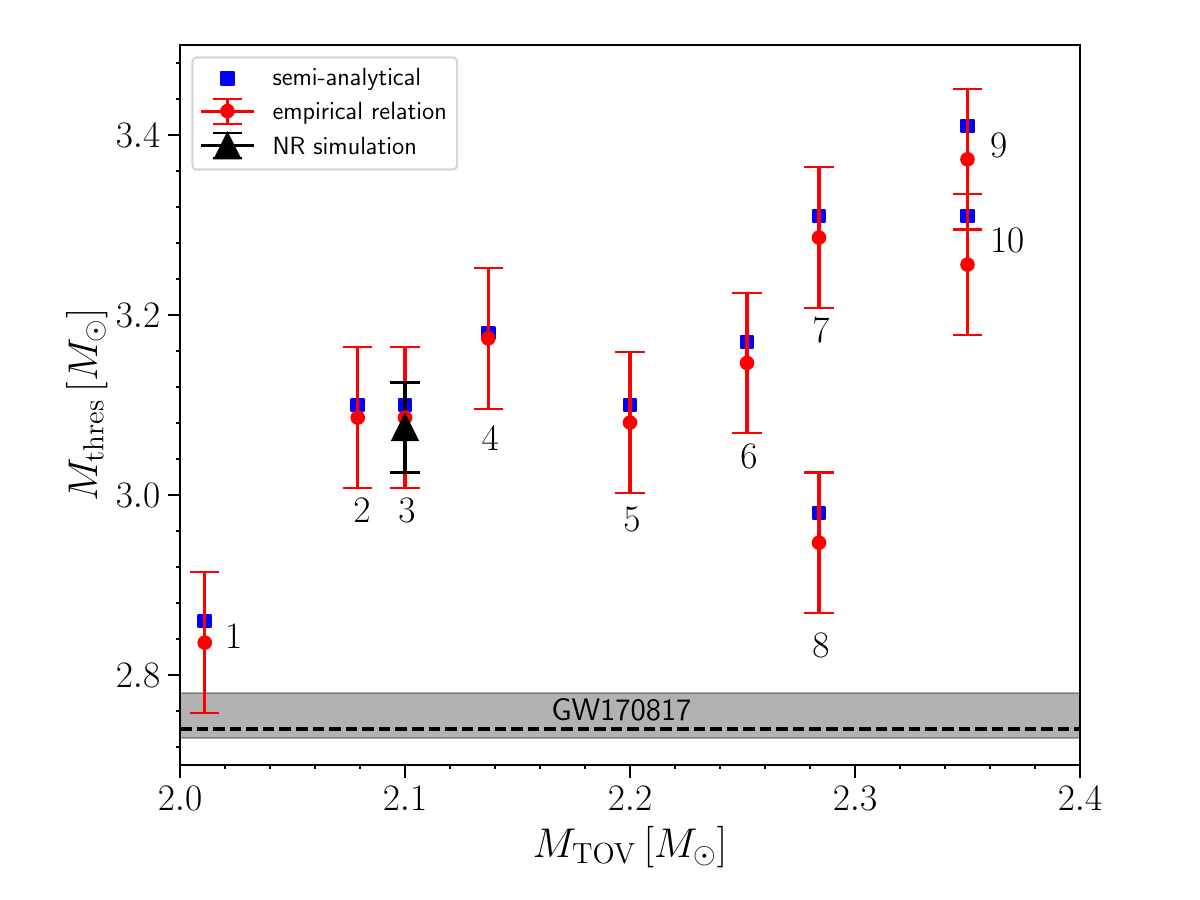}
   
	\end{center}
	\caption{$M_{\mathrm{thres}}$ for different QS EOSs obtained by different methods. The blue squares are $M_{\mathrm{thres}}$ calculated by the semi-analytical method with the $j$-constant law with $\hat{A}=2.0$. The red circles represent $M_{\mathrm{thres}}$ calculated by the empirical relation Eq. (\ref{eq:UR}). The red error bars indicate the maximum deviation between Eq. (\ref{eq:UR}) and the results of NR simulations of the BNS merger. The black triangle is $M_{\mathrm{thres}}$ calculated by the NR simulations and the black error bar indicates the uncertainty range of $M_{\mathrm{thres}}$ obtained by NR simulations due to the finite number of simulations. The numbers attached in
    each data denote the EOSs listed in Table~\ref{tab:eos}. The black horizontal dashed line represents the total binary
    ADM mass of GW170817 with the error bar indicated by the grey-dashed area.
   }
	\label{fig:thres}
\end{figure}

Except for EOS-3, there have been no current NR simulations to investigate $M_{\mathrm{thres}}$ for other QS EOSs. Here we calculate $M_{\mathrm{thres}}$ for QS EOSs listed in Table~\ref{tab:eos} through the semi-analytical method and derive the EOS-insensitive relation between $M_{\mathrm{thres}}$ and stellar properties.

Similar to $M_{\mathrm{TOV}}$ and $M_{\mathrm{thres}}$, the radius of the non-rotating maximum mass QS ($R_{\mathrm{TOV}}$) can also scale with the bag constant. Therefore, $M_{\mathrm{thres}}$ can be fitted with $M_{\mathrm{TOV}}$ and $R_{\mathrm{TOV}}$, and a bag constant independent relation can be obtained. The empirical relation among these quantities for NSs has been derived in~\citep{Bauswein2021} (No.18 fit) as:
\begin{equation}
   M_{\mathrm{thres}}=0.497 M_{\mathrm{TOV}}+0.1789 R_{\mathrm{TOV}}-0.004,
   \label{eq:UR}
\end{equation}
where $M_{\mathrm{thres}}$ and $M_{\mathrm{TOV}}$ are in units of $M_{\odot}$ and $R_{\mathrm{TOV}}$ is in units of km. The maximum deviation between this fit and $M_{\mathrm{thres}}$ obtained by NR simulations is $ 0.078\,M_{\odot}$. Employing QS EOSs listed in Table \ref{tab:eos}, we calculate $M_{\mathrm{thres}}$ for QSs through the empirical relation Eq. (\ref{eq:UR}) and the semi-analytical method, and the results are demonstrated in Figure~\ref{fig:thres}.

As shown in Figure~\ref{fig:thres}, it can be seen that $M_{\mathrm{thres}}$ obtained through the empirical relation Eq. (\ref{eq:UR}) and the semi-analytical method agree well. Although the last term in Eq. (\ref{eq:UR}) is a constant that can not scale with the bag constant, it is so small that it can be neglected. Therefore, the empirical relation Eq. (\ref{eq:UR}) is also valid for the BQS merger and can be used to constrain QS EOSs. With the assumption of no prompt collapse in GW170817~\citep{Bauswein2017,Margalit2017,Kasen2017}, the constraint on the parameter space of QS EOSs can be established and the results are plotted in Figure \ref{fig:parameter}. In addition, it can also be seen in Figure~\ref{fig:thres} that $M_{\mathrm{thres}}$ of the selected QS EOSs are larger than the total binary ADM mass of GW170817, and hence prompt collapse is unlikely to take place for these EOSs. The specific collapse behavior of the merger remnant with these EOSs will be investigated in Sec.~\ref{sec:170817}.

\section{DELAYED COLLAPSE}
\label{sec:delayed}
The merger of BNS with $M<M_{\mathrm{thres}}$ leads to the formation of a massive NS. If the mass of the remnant NS is greater than the maximum stable mass of uniformly rotating NSs, it has to be supported by differential rotation and is called a hypermassive NS (HMNS) \citep{Baumgarte2000,Kastaun2015,Hanauske2017}. For the case where the mass of the remnant NS is in the range of $M_{\mathrm{TOV}}$ to $M_{\mathrm{max,urot}}$, the remnant NS is called a
supramssive NS (SMNS) \citep{Cook1994,Lasota1996}. Both the HMNS and the SMNS are transiently stable NS and eventually undergo delayed collapse to BH due to the angular momentum redistribution and dissipation. The collapse of the HMNS remnant could be triggered by the angular momentum loss due to gravitational waves or through the angular momentum transport process in a timescale of $\sim 100 \,$ms~\cite{Kiuchi:2017zzg,Fujibayashi2018,Fujibayashi2020}, while the SMNS remnant can live in a spin-down timescale which is much longer than the HMNS case. 

For the BQS case, NR simulations in \citep{Zhou2022} have indicated that the gap between  
$M_{\mathrm{thres}}$ and $M_{\mathrm{max,urot}}$ is much smaller than the BNS case, which may lead to differences in the scenario of the post-merger product. Specifically, a supramassive remnant of the BQS merger could collapse to a BH within several times the dynamic timescale if the angular momentum contained in the remnant QS is not large enough to reach the mass-shedding limit~\citep{Zhou2022}. Thus, analyzing the angular momentum dissipation process is necessary for understanding the evolution of the merger remnant for the BQS merger. In this section, we consider the angular momentum dissipation process in the post-merger stage and employ both energy and angular momentum conservation laws to investigate the post-merger outcome for the BQS case.

\subsection{Equilibrium models of rotating QSs}
\label{sec:rotation}
In the delayed collapse case, an at least transiently stable QS is formed after the merger. We determine the stability and the stellar properties of the remnant rotating QS by constructing the equilibrium models. However, the rotation profile of the remnant QS is uncertain, because the remnant QS changes from a differential rotation state to a rigid rotation state due to the angular momentum redistribution process \citep{Shibata2017b,Fujibayashi2018}. For the merger remnant of GW170817, at the onset of collapse to a BH, whether a rigid rotation state was reached or not is still unclear. Nevertheless, our constraints of $M_{\mathrm{TOV}}$ just based on the exclusion of the very long-lived ($\sim 100\,\mathrm{s}$)/stable remnant that injects a large amount of its kinetic energy into the outflows from the observation results of the GW170817 event~\citep{Margalit2017, Shibata2017,Shibata2019}, and such remnant would achieve a rigid rotation state ($\sim 100\,$ms~\citep{Kiuchi:2017zzg,Fujibayashi2018,Fujibayashi2020}) before collapse to a BH. Therefore, to exclude QS EOSs with which the merger remnant is very long-lived or stable, it is enough to just focus on the uniform rotating case. In the rest of this paper, we employ the uniform rotation profile to construct the axisymmetric equilibrium models for the rotating QSs through the \texttt{RNS} code \citep{Stergioulas1995} and derive the relation between $M_{\mathrm{stab}}$ and $J$ by the turning point criterion.

\begin{figure}
	\begin{center}
		\includegraphics[height=70mm]{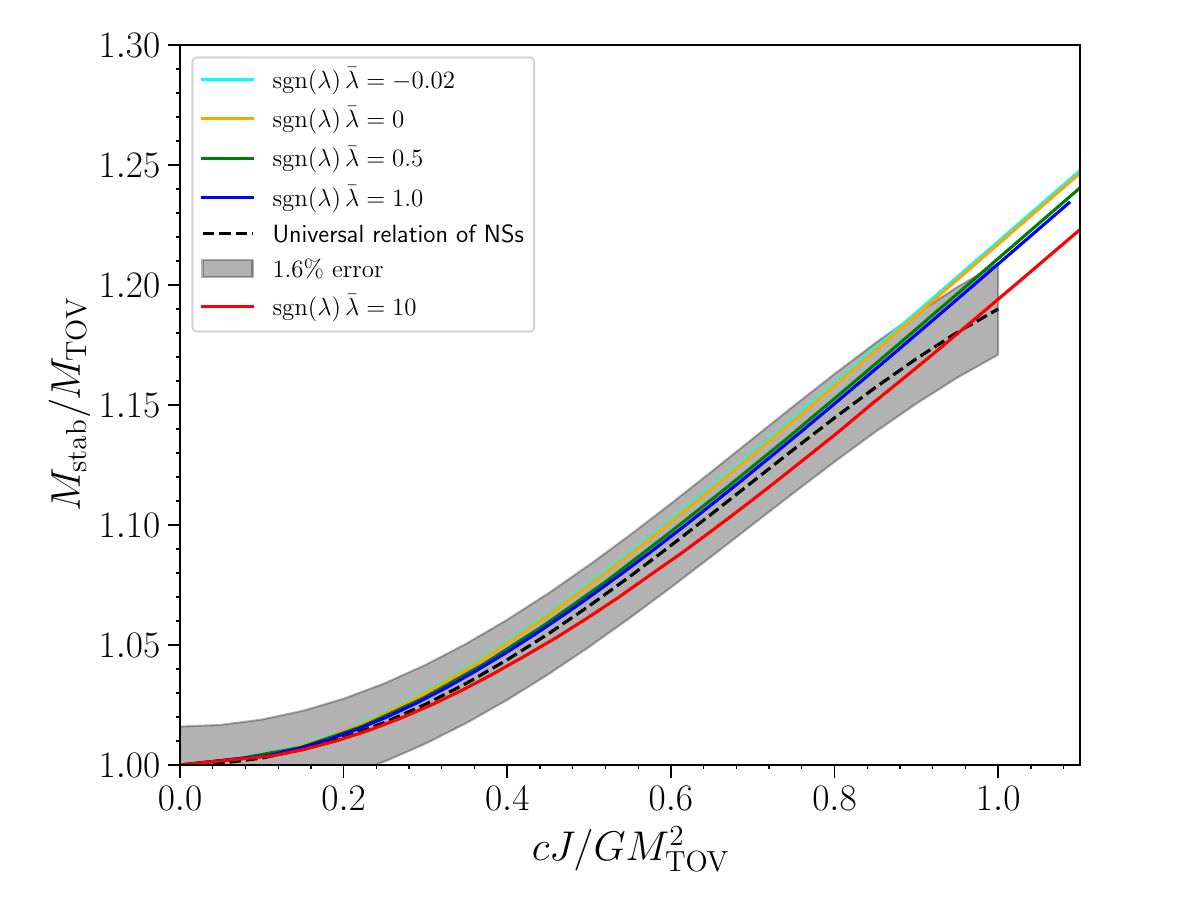}
	\end{center}
	\caption{The relation between the scaled maximum stable ADM mass and the scaled angular momentum of uniformly rotating QSs and NSs. Solid lines of different colors correspond to various values of $ \mathrm{sgn}(\lambda)\,\bar{\lambda} $, an EOS parameter of QSs. The black dashed line represents the universal relation of NSs in Ref. \citep{Bozzola2018}, where all the EOSs used in the study are within a $1.6\%$ error bar.}
	\label{fig:M-J}
\end{figure}

Figure \ref{fig:M-J} displays the relation between the scaled maximum stable mass $ \left(M_{\mathrm{stab}}/M_{\mathrm{TOV}} \right)$ and the scaled angular momentum $\left(c\, J/G\, M_{\mathrm{TOV}}^2 \right)$ for uniformly rotating QSs and NSs. The mass and angular momentum for QSs can be rescaled by $B_{\mathrm{eff}}$, and hence the relation is solely dependent on $\mathrm{sgn}(\lambda)\,\bar{\lambda}$. Figure \ref{fig:M-J} demonstrates that as $\mathrm{sgn}(\lambda)\,\bar{\lambda}$ increases, the scaled maximum stable mass decreases, given a scaled angular momentum. This is because the QS with larger $\mathrm{sgn}(\lambda)\,\bar{\lambda}$ is more compact (see Figure 1 in~\citep{Zhang2021}), and thus the gravitational field inside the star is stronger. Therefore, stronger centrifugal force is needed to sustain the additional self-gravity of the supramassive QS. In addition, it can be seen in Figure \ref{fig:M-J} that the $M_{\mathrm{stab}}-J$ relation is similar between NSs and QSs. Therefore, the value of $M_{\mathrm{max,urot}}/M_{\mathrm{TOV}}$ is larger for QSs mainly because at the mass-shedding limit more angular momentum is contained in QSs, but for a given $J$, the difference in $M_{\mathrm{stab}}$ is not significant (within $2\%$).

\begin{figure}[]
	\begin{center}
		\includegraphics[height=70mm]{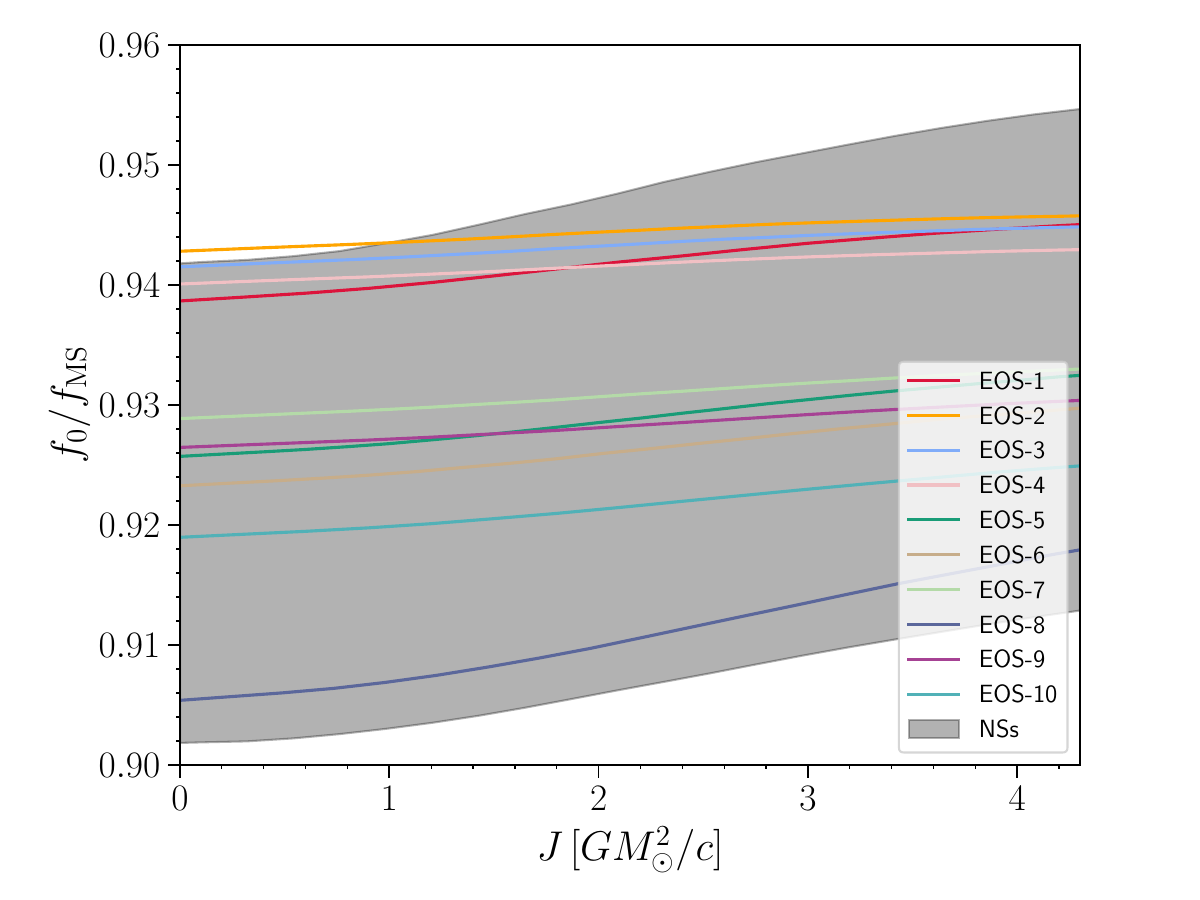}
	\end{center}
	\caption{The relation between $f_0/f_{\mathrm{MS}}$ and $J$ of uniformly rotating QSs, where $f_0$ is the ratio of the baryonic mass to the gravitational mass for binaries of masses $1.37\,M_{\odot}$ and $1.37\,M_{\odot}$, and $f_{\mathrm{MS}}$ represents the ratio of the baryonic mass to the gravitational mass at the turning points (a marginally stable state) for constant angular momentum sequences for uniformly rotating QSs. Different
    EOSs are shown with different colors. The grey-shaded region is the same relation for NSs with various piecewise polytrope EOSs \citep{Read2009} by which $2.0\,M_{\odot}<M_{\mathrm{TOV}}<2.4\,M_{\odot}$ and $\Lambda_{1.35}<1000$. 
   }
	\label{fig:f-J}
\end{figure}

\begin{figure}[]
	\begin{center}
		\includegraphics[height=70mm]{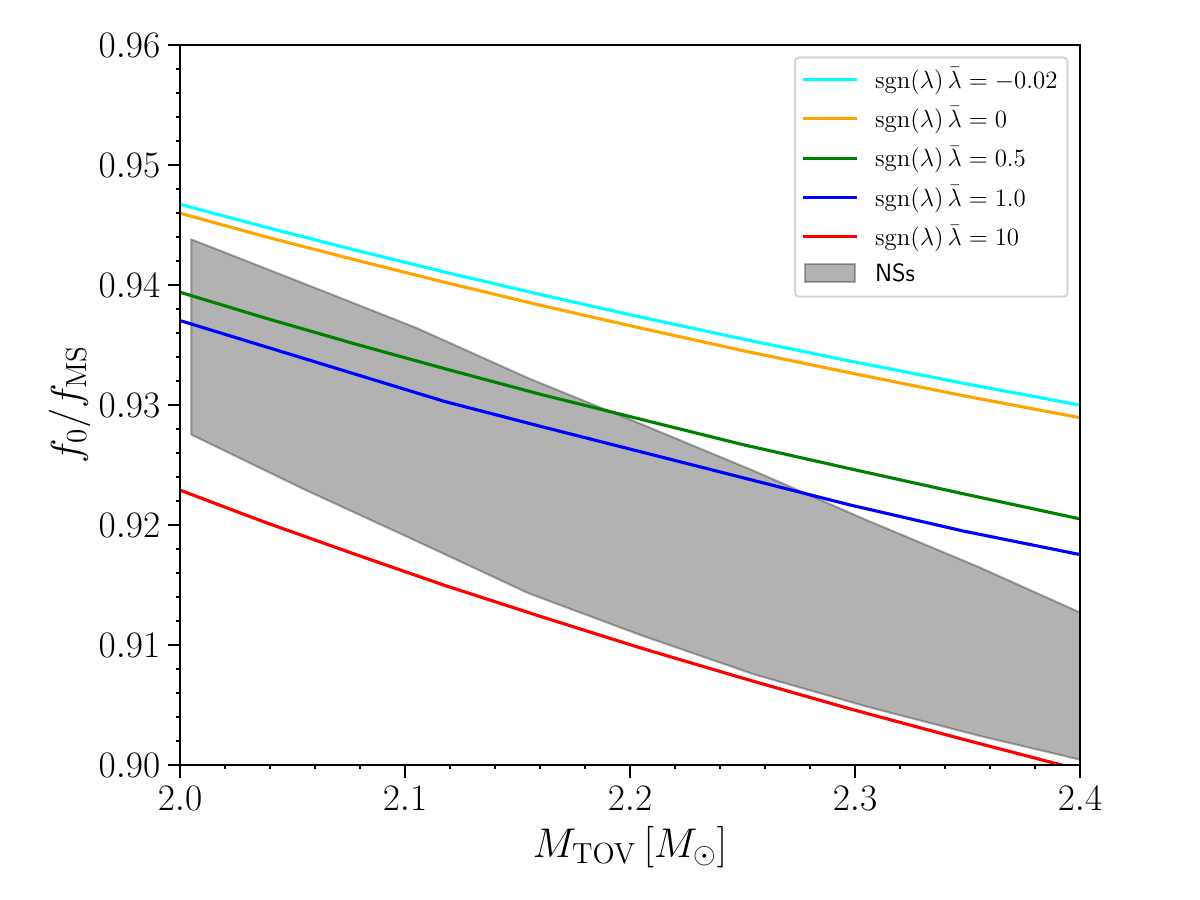}
	\end{center}
	\caption{The relation between $f_0/f_{\mathrm{MS}}$ and $M_{\mathrm{TOV}}$ for non-rotating QSs, where $f_0$ is the ratio of the baryonic mass to the gravitational mass for binaries of masses $1.37\,M_{\odot}$ and $1.37\,M_{\odot}$, and $f_{\mathrm{MS}}$ represents the ratio of the baryonic mass to the gravitational mass for marginally stable stars. Solid lines correspond to the relation between $f_0/f_{\mathrm{MS}}$ and $M_{\mathrm{TOV}}$ for EOSs with the same $ \mathrm{sgn}(\lambda)\,\bar{\lambda} $ but different $B_{\mathrm{eff}}$. Different colors correspond to various values of $ \mathrm{sgn}(\lambda)\,\bar{\lambda} $.  The grey-shaded region is the same relation for NSs with various piecewise polytrope EOSs \citep{Read2009} by which $2.0\,M_{\odot}<M_{\mathrm{TOV}}<2.4\,M_{\odot}$ and $\Lambda_{1.35}<1000$. 
 }
	\label{fig:f0}
\end{figure}

In the delayed collapse case, a QS surrounded by a torus is formed after the merger, and the mass ejection processes would occur. The baryonic masses ($M_{\ast}$)  of the remnant QS, the torus and the ejecta can be related through the baryon number conservation equation. From the time of the merger to the onset of the remnant QS collapse to a BH, the baryon number conservation equation can be expressed as:

\begin{equation}
    M_{\ast f}=M_{\ast}-M_{\ast\mathrm{out}}-M_{\ast\mathrm{eje}},  \label{5} 
\end{equation}
where $M_{\ast f}$ represents the baryonic mass of the remnant QS at the onset of collapse to a BH, $M_{\ast}$ denotes the baryonic mass of the binary system at its formation, $M_{\ast\mathrm{out}}$ is the baryonic mass of the torus surrounding the remnant BH at its formation, $M_{\ast\mathrm{eje}}$ corresponds to the baryonic mass of the ejecta at the time of BH formation. Here we conventionally set the baryonic mass per baryon number for the torus, the ejecta and the QS equal to the atomic mass unit ($m_{u}=931.494\,\mathrm{MeV}/c^2$) and the baryon number conservation can be written as the baryonic mass conservation.

To relate the baryonic mass to the gravitational mass, we define $f_0:=M_{\ast}/M$ and $f_{\mathrm{MS}}:=M_{\ast f}/M_{f}$. According to \citep{Friedman1988}, a secular instability occurs at the turning point (a marginally stable state) of the $M-\rho_{\mathrm{c}}$ curve with constant $J$. Hence, $f_{\mathrm{MS}}$ can be obtained by calculating the ratio of the baryonic mass to the gravitational mass at the turning point, and the baryon number conservation equation can be further expressed as:
\begin{equation}
    \begin{split}
         M_{\ast\mathrm{out}}+M_{\ast\mathrm{eje}}\approx&\,0.11\,M_{\odot}\left(\frac{f_{\mathrm{MS}}}{1.2}\right)\\
         &\times \left[\left(\frac{f_0/f_{\mathrm{MS}}}{0.925}\right)\left(\frac{M}{2.8\,M_{\odot}}\right)-\left(\frac{M_f}{2.5\,M_{\odot}}\right)\right]. \label{9}
    \end{split}
\end{equation}
From Eq. (\ref{9}), it can be found that $f_0/f_{\mathrm{MS}}$ deeply affects the sum of the masses of the torus and the ejecta. One percent increase of $f_0/f_{\mathrm{MS}}$ results in $\sim 0.02\,M_{\odot}$ increase in the sum of the masses of the torus and the ejecta, for the fixed $M_f$ and $M$.

Figure \ref{fig:f-J} shows the relation between $f_0/f_{\mathrm{MS}}$ and $J$ for QSs and NSs, where $f_0$ is the ratio of the baryonic mass
to the gravitational mass for binaries of masses $1.37\,M_{\odot}$ and $1.37\,M_{\odot}$ inferred from GW170817 with the symmetric binary assumption \footnote{It is known that for QSs described by IQM models, their masses scale with $1/\sqrt{B_\mathrm{eff}}$. In other words, for different $B_\mathrm{eff}$, $f=M_*/M$ remains unchanged if the masses of the model are multiplied by $\sqrt{B_\mathrm{eff}}$. Therefore, $f_\mathrm{MS}$ is independent of $B_\mathrm{eff}$ as we are considering the TOV maximum mass solution, and the $M_\mathrm{TOV}$ of QSs indeed scales as $1/\sqrt{B_\mathrm{eff}}$. However, $f_0$ is dependent on  $B_\mathrm{eff}$, since we are considering models with a fixed mass of $1.37\,M_\odot$, instead of a scaled mass.}. It can be seen that this relation is similar for NSs and QSs, and thus the relation among $M_{\ast\mathrm{out}}+M_{\ast\mathrm{eje}}$, $M$ and $M_f$ is also similar for NSs and QSs. Moreover, we fix $\mathrm{sgn}(\lambda)\,\bar{\lambda}$ and change $B_{\mathrm{eff}}$ to derive the relation between $f_0/f_{\mathrm{MS}}$  and $M_{\mathrm{TOV}}$ for non-rotating QSs, for which $f_{\mathrm{MS}}=f_{\mathrm{TOV}}$ because the turning point for the non-rotating case corresponds to the maximum mass spherical configuration. As shown in Figure \ref{fig:f0}, the increases of $M_{\mathrm{TOV}}$ and $\mathrm{sgn}(\lambda)\,\bar{\lambda}$ lead to the decrease of $f_0/f_{\mathrm{MS}}$. Therefore, for the fixed $M_{\mathrm{TOV}}$, $M$ and $M_f$, the value of $M_{\ast\mathrm{out}}+M_{\ast\mathrm{eje}}$ for QS EOSs with larger $\mathrm{sgn}(\lambda)\,\bar{\lambda}$ would be smaller. This trend matches the expectation from a simple qualitative picture: the one-to-one positive correlation between $\bar{M}_{\text{TOV}}$ (defined as $M_{\text{TOV}}\sqrt{4B_{\text{eff}}}$) and $\bar{\lambda}$~\citep{Zhang2021} indicates that a larger $\text{sgn}(\lambda)\bar{\lambda}$ for a given $M_{\text{TOV}}$ maps to a larger bag constant, which characterizes the inward pressure from the QCD vacuum differences that render the QSs more bounded and thus smaller $M_{\ast\text{out}}+M_{\ast\text{eje}}$.

\subsection{Dissipation mechanisms and conservation equations}
\label{sec:dissipation}

After the merger, the remnant NS is highly deformed and has a non-axisymmetric configuration, and hence quasi-periodic gravitational waves are emitted from the remnant NS \citep{Bauswein2012,Bauswein2012b,takami2014, Shibata2006, Kiuchi2009,  Sekiguchi2011, Sekiguchi2011b,Bernuzzi2015,Oechslin2007,Stergioulas2011,Hotokazaka2013,Takami2015,Dietrich2017}. Copious neutrinos are also emitted from the remnant NS because, after the merger, strong shocks are generated and heat the remnant NS \citep{Eichler1989,Ruffert1997,Ruffert2001,Rosswog2003,Sekiguchi2015,Palenzuela2015,Foucart2016,Foucart2016b,Wu2017,George2020,Kullmann2022,Cusinato2021,Radice2022,Espino2024,Fujibayashi2018,Fujibayashi2020}.  Moreover, a fraction of the matter obtains sufficient specific angular momentum to result in a torus or ejecta through the angular momentum transport process \citep{Hotokezaka2013,Hotokezaka2013b,Fujibayashi2018,Fujibayashi2020,Kiuchi:2022nin}. Although electromagnetic radiation like magnetic dipole radiation also dissipates the energy and the angular momentum of the remnant NS, it only has a minor effect unless the remnant is very long-lived ($\sim 100\, \mathrm{s}$). Such a very long-lived remnant would have injected a large fraction of its total rotational kinetic energy ($\sim 10^{53}\,$erg) into the ejecta, which is disfavored by the observational results of GW170817~\citep{Margalit2017,Shibata2019}, and thus the dissipation due to electromagnetic radiation is not taken into account in this paper. Similar to the BNS merger, these dissipation processes also take place in the BQS scenario \citep{Bauswein2009,Bauswein2010,Zhou2021,Zhou2024}. Therefore, the energy and the angular momentum conservation equations for the BQS merger from the moment of the merger to the onset of the collapse can be expressed as:

\begin{gather}
    M_f=M_{\mathrm{merger}}-E_{\mathrm{GW,p}}-E_{\nu}-M_{\mathrm{out}}-M_{\mathrm{eje}}, \label{4} \\
    J_f=J_{\mathrm{merger}}-J_{\mathrm{GW,p}}-J_{\nu}-J_{\mathrm{out}}-J_{\mathrm{eje}}, \label{6}
\end{gather}
where $M_f$ and $J_f$ are the gravitational
mass and angular momentum of the remnant QS at the onset of the collapse to a BH, respectively; $M_{\mathrm{merger}}$ and $J_{\mathrm{merger}}$ denote the gravitational mass and angular momentum at the time of the merger; $E_{\mathrm{GW,p}}$ and $J_{\mathrm{GW,p}}$ represent the energy and angular momentum carried away by gravitational waves in the post-merger phase; $E_{\mathrm{\nu}}$ and $J_{\mathrm{\nu}}$ are the energy and angular momentum dissipated through neutrino emission; $M_{\mathrm{out}}$ and  $J_{{\mathrm{out}}}$ correspond to the gravitational mass and angular momentum of the torus surrounding the remnant BH at its formation; $M_{\mathrm{eje}}$ and  $J_{{\mathrm{eje}}}$ are the gravitational mass and angular momentum of the ejecta at the time of BH formation. In this paper, it is assumed that the internal, kinetic, and gravitational binding energy of the matter outside the merger remnant is significantly smaller compared to its baryonic mass, and thus we do not distinguish the baryonic mass and the gravitational mass of the matter outside the merger remnant (i.e., $M_{\ast\mathrm{out}}=M_{\mathrm{out}}$ and $M_{\ast{\mathrm{eje}}}=M_{\mathrm{eje}}$).

To solve the baryon number, the energy and the angular momentum conservation equations in the post-merger phase, we first need to calculate the baryonic mass, the gravitational mass and the angular momentum of the remnant QS at the time of the merger which serve as the initial properties of the remnant QS. The baryon number is conserved in the inspiral phase, and hence we can calculate the baryonic mass of the binary at the time of merger by solving the TOV equation, given the mass components and the EOS. $J_{\mathrm{merger}}$ can be accurately computed by the EOB method as shown in Sec.~\ref{sec:prompt}. Similar to $J_{\mathrm{merger}}$, $M_{\mathrm{merger}}$ can also be obtained via the EOB method. The energy carried away by gravitational waves during the inspiral phase ($E_{\mathrm{GW,i}}$) can be calculated via an integration of the gravitational waveform generated by the EOB method:
\begin{equation}
    E_{\mathrm{GW,i}}=\frac{1}{16\pi} \sum\limits_{(\ell,m)}\int_{t_0}^{t_{\mathrm{merger}}}dt^{\prime} r^2\dot h_{\ell m}\left( t^{\prime} \right) \dot h_{\ell m}^* \left( t^{\prime} \right).\label{7}
\end{equation}
Therefore, $M_{\mathrm{merger}}$ can be expressed as:
\begin{equation}   
    M_{\mathrm{merger}}=M_0-E_{\mathrm{GW,i}} ,
\end{equation}
where $M_0$ is the energy of the binary at the beginning of the calculation of gravitational wave, and can be derived through Eqs. (2.4) and (4.3) in \citep{Buonanno2000}.

At the onset of collapse, the configuration of the remnant QS can be estimated by the equilibrium configuration at the turning point of the $M-\rho_{\mathrm{c}}$ curve with constant $J$, where the secular instability occurs \citep{Friedman1988}. Therefore, the baryonic mass, the gravitational mass and the angular momentum of the remnant QS at the onset of collapse, which serve as the final properties of the remnant QS, can be calculated as long as one of these three quantities is given.

As discussed above, given $M_f$ (or $J_f$ and $M_{\ast f}$) we can obtain the dissipated baryonic mass, gravitational mass and angular momentum. Furthermore, there are three main dissipation mechanisms, the gravitational radiation, the neutrino emission and the hydrodynamical angular momentum transport process, which leads to the formation of the torus and the ejecta, and obviously, they dissipate the baryonic mass, the energy and the angular momentum in different proportions. Therefore, in principle, we can determine the amount of dissipation of the three dissipation mechanisms via the three conservation equations, which is equivalent to solving a linear system with three unknowns. However, before solving this linear system, we should know the proportion of the dissipated baryonic mass, energy and angular momentum for each mechanism, which serves as the coefficient of each unknown and has been obtained through NR simulations as summarized in \citep{Shibata2019}. Here we briefly introduce them.

For gravitational waves in the post-merger phase, the peak frequency, $f_{\mathrm{peak}}$, in the spectrum amplitude corresponds to the frequency of the $f$–mode oscillation of the remnant massive NS \citep{Bauswein2012,Hotokezaka2013,Rezzolla2016}, and it is reported to correlate with the tidal deformability \citep{Rezzolla2016, Kiuchi2020}, and the approximate expression for $J_{\mathrm{GW,p}}$ can be expressed as:
\begin{equation}
    \begin{split}
        J_{\mathrm{GW,p}}&\approx \frac{E_{\mathrm{GW,p}}}{\pi f_{\mathrm{peak}}} \\
                &\approx 9.5\times 10^{48} \,\mathrm{erg\,s} \left(\frac{E_{\mathrm{GW,p}}}{0.05\,M_{\odot}}\right)\left(\frac{f_{\mathrm{peak}}}{3.0\,\mathrm{kHz}}\right)^{-1} . \label{10}
    \end{split}
\end{equation}
This relation has been confirmed by NR simulations \citep{Kiuchi2020}, with a relative error $\lesssim 1\%$. The value of $f_{\mathrm{peak}}$ can be determined using the relation provided in Eq. (5.4) of Ref. \citep{Kiuchi2020}. Furthermore, this relation is consistent with the BQS merger simulations in \citep{Zhou2024}, and hence we employ this relation for the BQS scenario.

For neutrino emission, as the emitter (remnant QS) undergoes rotation, the neutrinos carry away the angular momentum of the remnant QS. Therefore, $J_{\nu}$ can be approximately written as $J_{\nu}\approx(2/3)R_{\mathrm{MQS}}^2\Omega E_{\nu}$ \citep{Baumgarte1998}, and correspondingly
\begin{equation}
    \begin{split}
        J_{\nu} \approx & 3.0\times 10^{48} \,\mathrm{erg\,s} \left( \frac{E_{\nu}}{0.1\,M_{\odot}c^2}\right) \left( \frac{R_{\mathrm{MQS}}}{15\, \mathrm{km}} \right)^2 \\
         & \times\left( \frac{\Omega}{10^4 \, \mathrm{rad/s}}\right),
         \label{11}
    \end{split}
\end{equation}
where $R_{\mathrm{MQS}}$ and $\Omega$ represent the equatorial circumferential radius and angular velocity of the remnant massive QS. This relation is consistent with the NR simulation results within a factor of 2 \citep{Fujibayashi2020}. In addition, comparing Eq. (\ref{11}) with Eq. (\ref{10}), we find that neutrino emission serves as a secondary mechanism for the dissipation of angular momentum, and thus the uncertainty in neutrino emission only has a minor impact on the results.

The velocity of the torus surrounding the remnant QS and the ejecta should be of the order of their escape velocity, and thus the angular momentum carried away by them can be approximately expressed as $J_{\mathrm{out}}\approx M_{\mathrm{out}}\sqrt{GM_{\mathrm{MQS}}R_{\mathrm{out}}}$ and $ J_{\mathrm{eje}}\approx M_{\mathrm{eje}}\sqrt{GM_{\mathrm{MQS}}R_{\mathrm{eje}}} $, i.e.,

\begin{equation}
    \begin{split}
        J_{\mathrm{out}} \approx &5.8\times 10^{48} \,\mathrm{erg\,s}  \left( \frac {M_{\mathrm{out}}}{0.05\,M_{\odot}}\right) \left( \frac{R_{\mathrm{out}}}{100\, \mathrm{km}} \right)^{1/2} \\
        & \times\left( \frac{M_{\mathrm{MQS}}}{2.6 \, M_{\mathrm{\odot}}}\right)^{1/2},
         \label{12}
    \end{split}
\end{equation}

\begin{equation}
    \begin{split}
        J_{\mathrm{eje}} \approx  &6.9\times 10^{48} \,\mathrm{erg\,s}  \left( \frac {M_{\mathrm{eje}}}{0.05\,M_{\odot}}\right) \left( \frac{R_{\mathrm{eje}}}{140\, \mathrm{km}} \right)^{1/2} \\
        & \times\left( \frac{M_{\mathrm{MQS}}}{2.6 \, M_{\mathrm{\odot}}}\right)^{1/2} ,
         \label{13}
    \end{split}
\end{equation}
where ${M_{\mathrm{MQS}}}$ represents the gravitational mass of the remnant massive QS, $R_{\mathrm{out}}$ is the typical radius of the torus surrounding the remnant QS, and $R_{\mathrm{eje}}$ is the typical location where ejection occurs. Both gravitational mass and equatorial circumferential radius for the remnant QS at the onset of collapse can be estimated by constructing the equilibrium models with the turning point criterion for the given angular momentum or total baryonic mass. The NR simulations for the BNS merger have shown that $R_{\mathrm{out}}$ would experience an increase from $\sim 50\, \mathrm{km}$ to $\sim 200\, \mathrm{km}$ due to the long-term viscous evolution of the torus, as well as the transport of angular momentum from the remnant NS \citep{Fujibayashi2018,Kiuchi:2022nin}. It has been suggested that the dynamics of the BQS merger and the BNS merger are similar \citep{Zhou2024}. Therefore, the evolution of the torus may not make much difference and we use the same relation as in \citep{Shibata2019}, where it is simply set that $R_{\mathrm{out}}=R_{\mathrm{eje}}=140\,\mathrm{km}$. In addition, $J_{\mathrm{out}}$ mainly depends on $M_{\mathrm{out}}$ instead of $R_{\mathrm{out}}$ ($J_{\mathrm{out}}\propto M_{\mathrm{out}}$, while $J_{\mathrm{out}}\propto \sqrt{R_{\mathrm{out}}}$), and thus the value of $R_{\mathrm{out}}$ and $R_{\mathrm{eje}}$ do not make a great effect on our analysis.

\section{APPLICATION TO GW170817 EVENT}
\label{sec:170817}
As mentioned in Sec.~\ref{sec:introduction}, for the GW170817 event, the prompt collapse is not likely to take place due to the very presence of an electromagnetic counterpart \citep{Margalit2017,Kasen2017,Bauswein2017}\footnote{Although it is suggested in \citep{Kiuchi2019} that the luminosity of AT 2017gfo could be explained by models that undergo prompt collapse, blue kilonova component of AT 2017gfo also disfavors the case of prompt collapse as discussed in these studies.}. Therefore, the constraint on the threshold mass for prompt collapse can be made $M_{\mathrm{thres}}\gtrsim2.74\,M_{\odot}$.
On the other hand, a very long-lived ($\sim 100\, \mathrm{s}$)/stable remnant that injects a large number of its kinetic energy into the outflows is also disfavored because, with such strong energy injection, the ultraviolet-optical-infrared counterparts for GW170817 would be much brighter than the observational results \citep{Margalit2017,Shibata2017,Shibata2019}\footnote{A long-lived NS as merger remnant for the GW170817 event can not be completely ruled out, and this scenario has been investigated in~\citep{Li2018,Ai2018,Piro2019,Dupont2024}. However, in this paper, our constraint on QS EOSs is only based on the predominant scenario proposed in~\citep{Margalit2017,Shibata2017,Shibata2019}.}. Therefore, our analysis in this Sec. \ref{sec:delayed} can be directly applied to the GW170817 event to understand the post-merger evolution of this event and set the constraint on QS EOSs.

\begin{table*}[]

    \centering
	\begin{tabular}{c@{\hskip 20pt}c@{\hskip 15pt}c@{\hskip 12pt}c@{\hskip 17pt}c@{\hskip 22pt}c@{\hskip 22pt}c@{\hskip 22pt}c@{\hskip 22pt}c@{\hskip 7pt}c}
		\hline
		\hline
        Model & $M_{\mathrm{merger}}$ & $J_{\mathrm{merger}}$  & $M_f/M_{\mathrm{TOV}}$ & $M_f$  & $J_f$  & $E_{\mathrm{GW,p}}$ &$E_{\mathrm{GW}}$  & $E_{\nu}$ & $M_{\mathrm{out}}+M_{\mathrm{eje}}$   \\
        \hline
        \multirow {5}{*}{EOS-1} & \multirow {5}{*}{2.695} & \multirow {5}{*}{6.192} & 1.269 & 2.550 & 4.812 & 0.022 & 0.067 & 0.073 & 0.05  \\
                                                                                &  &  & ... & ... & ... & ...&... & ... & 0.10\\
                                                                                &  &  & ... & ... & ... & ...&... &... & 0.15 \\&  &  & ... & ... & ... & ...&... &... & 0.20 \\
                                                                                
        \hline
        \multirow {5}{*}{EOS-2} & \multirow {5}{*}{2.700} & \multirow {5}{*}{6.438} & 1.226 & 2.550 & 4.434 & 0.046 & 0.086 & 0.054 & 0.05  \\
                                &                         &                         & 1.206 & 2.508
        & 4.139 & 0.025& 0.065 & 0.067 & 0.10 \\
                                &                         &                         & 1.186 & 2.466
        & 3.844 & 0.006 & 0.046 & 0.078 & 0.15 \\ 
         &                         &                         & 1.186 & 2.466
        & 3.844 & 0.006 & 0.046 & 0.078 & 0.15 \\  &                         &                         & ... & ...
        & ... & ... & ... & ... & 0.20 \\ 
        \hline
        \multirow {5}{*}{EOS-3} & \multirow {5}{*}{2.700} & \multirow {5}{*}{6.424} & 1.213 & 2.547 & 4.335 & 0.050 & 0.091 & 0.052 & 0.05  \\
                                &                         &                         & 1.193 & 2.505
        & 4.037 & 0.030 & 0.070 & 0.065 & 0.10 \\
                                &                         &                         & 1.173 & 2.463
        & 3.737 & 0.010 & 0.051 & 0.076 & 0.15 
        \\
        
                                &                         &                         & ... & ...
        & ... & ... & ... & ... & 0.20
        \\
        \hline
        \multirow {5}{*}{EOS-4} & \multirow {5}{*}{2.701} & \multirow {5}{*}{6.488} & 1.190 & 2.542 & 4.117 & 0.059 & 0.099 & 0.049 & 0.05  \\
                                &                         &                         & 1.170 & 2.501
        & 3.816 & 0.041 & 0.080 & 0.059 & 0.10 \\
                                &                         &                         & 1.151 & 2.459
        & 3.512 & 0.022 & 0.062 & 0.069 & 0.15 \\ 
                                &                         &                         & 1.131 & 2.418
        & 3.204 & 0.005 & 0.045 & 0.078 & 0.20 \\ 
        \hline
        \multirow {5}{*}{EOS-5} & \multirow {5}{*}{2.697} & \multirow {5}{*}{6.274} & 1.143 & 2.516 & 3.746 & 0.080 & 0.123 & 0.052 & 0.05  \\
                                &                         &                         & 1.126 & 2.477
        & 3.445 & 0.057 & 0.101 & 0.062 & 0.10 \\
                                &                         &                         & 1.108 & 2.439
        & 3.137 & 0.036 & 0.080 & 0.072 & 0.15 \\
                        &                         &                         & 1.091 & 2.400
        & 2.817 & 0.016 & 0.060 & 0.080 & 0.20 \\
        
        \hline
        \multirow {5}{*}{EOS-6} & \multirow {5}{*}{2.697} & \multirow {5}{*}{6.311} & 1.112 & 2.505 & 3.356 & 0.098 & 0.141 & 0.044 & 0.05  \\
                                &                         &                         & 1.094 & 2.464
        & 3.015 & 0.078 & 0.121 & 0.055 & 0.10 \\
                                &                         &                         & 1.076 & 2.424
        & 2.657 & 0.060 & 0.103 & 0.063 & 0.15 \\
         &                         &                         & 1.058 & 2.383
        & 2.274 & 0.044 & 0.087 & 0.070 & 0.20 \\
        \hline
        \multirow {5}{*}{EOS-7} & \multirow {5}{*}{2.701} & \multirow {5}{*}{6.478} & 1.102 & 2.516 & 3.164 & 0.100 & 0.140 & 0.034 & 0.05  \\
                                &                         &                         & 1.085 & 2.477
        & 2.832 & 0.083 & 0.123 & 0.040 & 0.10 \\
                                &                         &                         & 1.067 & 2.438
        & 2.479 & 0.067 & 0.107 & 0.045 & 0.15 \\
         &                         &                         & 1.050 & 2.399
        & 2.099 & 0.053 & 0.093 & 0.048 & 0.20 \\
        \hline
       
        \multirow {5}{*}{EOS-8} & \multirow {5}{*}{2.692} & \multirow {5}{*}{6.078} & 1.079 & 2.465 & 2.992 & 0.127 & 0.175 & 0.050 & 0.05  \\
                                &                         &                         & 1.060 & 2.422
        & 2.571 & 0.108 & 0.156 & 0.061 & 0.10 \\
                                &                         &                         & 1.042 & 2.379
        & 2.094 & 0.094 & 0.142 & 0.069 & 0.15 \\
        
                                &                         &                         & 1.023 & 2.336
        & 1.519 & 0.087 & 0.135 & 0.069 & 0.20 \\
        \hline
        \multirow {5}{*}{EOS-9} & \multirow {5}{*}{2.701} & \multirow {5}{*}{6.521} & 1.066 & 2.505 & 2.599 & 0.121 & 0.160 & 0.024 & 0.05  \\
                                &                         &                         & 1.049
        & 2.465 & 2.192 & 0.108 & 0.147 & 0.028 & 0.10 \\
                                &                         &                         & 1.032
        & 2.425 & 1.731 & 0.097 & 0.136 & 0.029 & 0.15 \\
        
                                &                         &                         & 1.015
        & 2.385 & 1.151 & 0.091 & 0.130 & 0.025 & 0.20 \\
        \hline
        \multirow {5}{*}{EOS-10} & \multirow {5}{*}{2.699} & \multirow {5}{*}{6.380} & 1.059 & 2.489 & 2.493 & 0.136 & 0.177 & 0.024 & 0.05  \\
                                &                         &                         & 1.042
        & 2.449 & 2.061 & 0.121 & 0.162 & 0.029 & 0.10 \\
                                &                         &                         & 1.024
        & 2.409 & 1.552 & 0.112 & 0.153 & 0.028 & 0.15 \\
        
                                &                         &                         & 1.008
        &2.368  & 0.829 & 0.119 & 0.160 & 0.018 & 0.20 \\
        \hline
		\hline
	\end{tabular}
	\caption{The properties ($M_f/M_{\mathrm{TOV}}$, $M_f$, $J_f$, $E_{\mathrm{GW,p}}$, $E_{\mathrm{GW}}$, $E_{\nu}$, $M_{\mathrm{out}}+M_{\mathrm{eje}}$) of the potential states of uniformly rotating QSs at the onset of collapse in the GW170817 event are shown, considering various EOSs. The units of the mass, energy and angular momentum are $M_{\odot}$, $M_{\odot}c^2$ and $GM_{\odot}^2/c$, respectively. The chosen values of $M_{\mathrm{out}}+M_{\mathrm{eje}}$ are $0.05\,M_{\odot}$, $0.10\,M_{\odot}$, $0.15\,M_{\odot}$, $0.20\,M_{\odot}$. The values of $M_{\mathrm{merger}}$ and $J_{\mathrm{merger}}$ are calculated  by EOB method. `...' means only solutions with negative values of $E_{\mathrm{GW,p}}$ are obtained.  }
	\label{tab:results}
    
\end{table*}

 We apply the analysis in Sec. \ref{sec:delayed} to the GW170817 event and search for solutions that satisfy the energy and angular momentum conservation laws for the selected QS EOSs. We fix the total gravitational mass of binary $M=2.74\,M_{\odot}$. The mass ratio $M_1/M_2$ only has little effect on the final results \citep{Shibata2019}, and thus we only show the results with $M_1=M_2=1.37\,M_{\odot}$. As discussed in \ref{sec:dissipation}, given a series of $M_f$, the values of $E_{\mathrm{GW,p}}$, $E_{\nu}$ and $M_{\mathrm{out}}+M_{\mathrm{eje}}$ can be calculated, and thus the relation between them can also be obtained. To determine whether or not the solutions are reasonable, we should check whether the obtained solution range could satisfy the observation results and the constraints placed by NR simulations. In this paper, we employ the NR simulation results in \citep{Zappa2018}, where the value of the total energy radiated by gravitational
waves throughout the inspiral to the post-merger phases ($E_{\mathrm{GW}}$) can be constrained as:
\begin{equation}
  E_{\mathrm{GW}}:= E_{\mathrm{GW,i}}+E_{\mathrm{GW,p}} \lesssim 0.13 \pm 0.01\, M_{\odot}c^2 \left( \frac{M}{2.8\,M_{\odot}} \right). \label{19}
\end{equation}

The uncertainty is due to the finite grid resolution and the finite sample size. Although this relation is obtained through NR simulations for BNS merger, due to the similarity between the dynamics of the BQS merger and the BNS merger \citep{Zhou2024}, we can employ this relation for the BQS merger. Only QS EOSs with which self-consistent solutions (i.e., three conservation equations could be solved with reasonable values of $E_{\mathrm{GW}}$, $E_\nu$ and $M_\mathrm{out}+M_\mathrm{eje}$) exist can pass our consistency checks.

\begin{figure*}[]
	\begin{center}
		\includegraphics[width=0.48\textwidth]{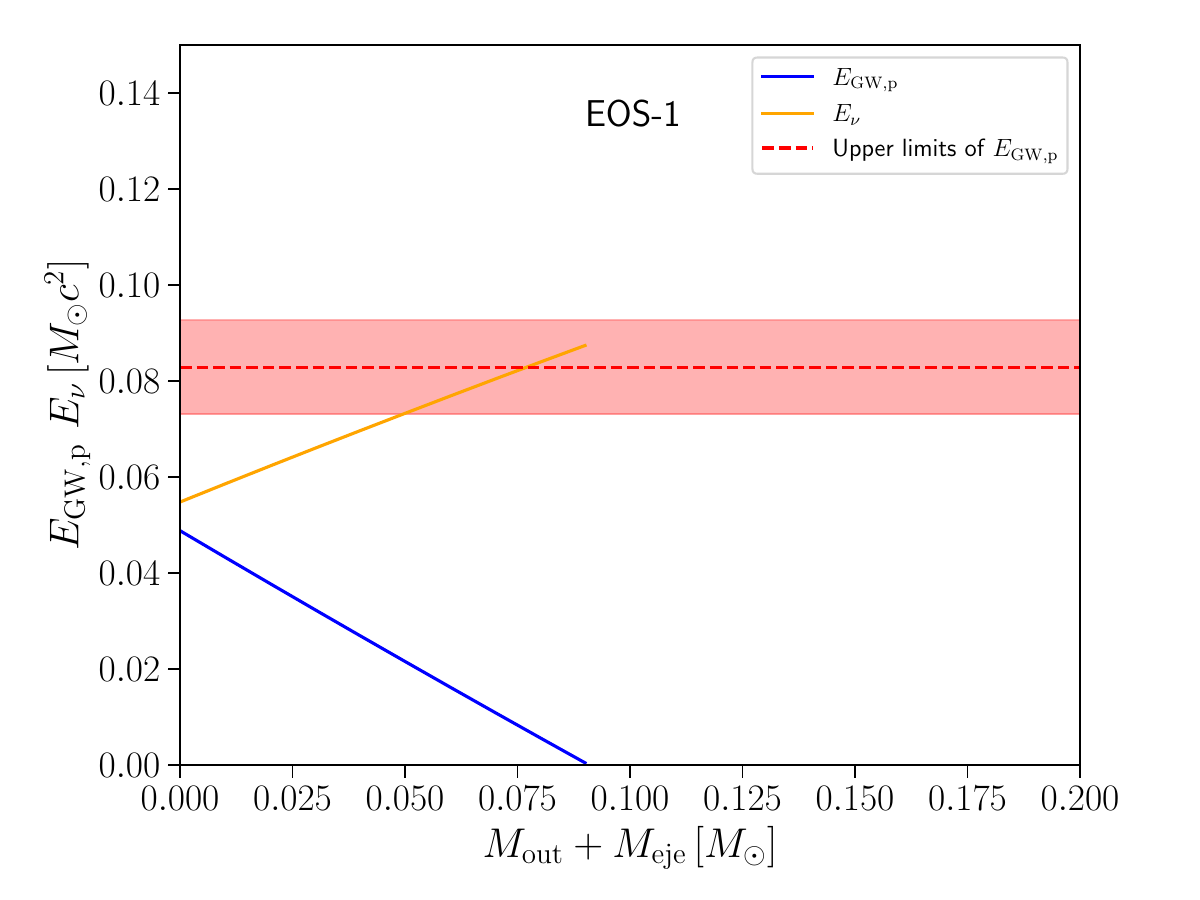}
        \quad
        \includegraphics[width=0.48\textwidth]{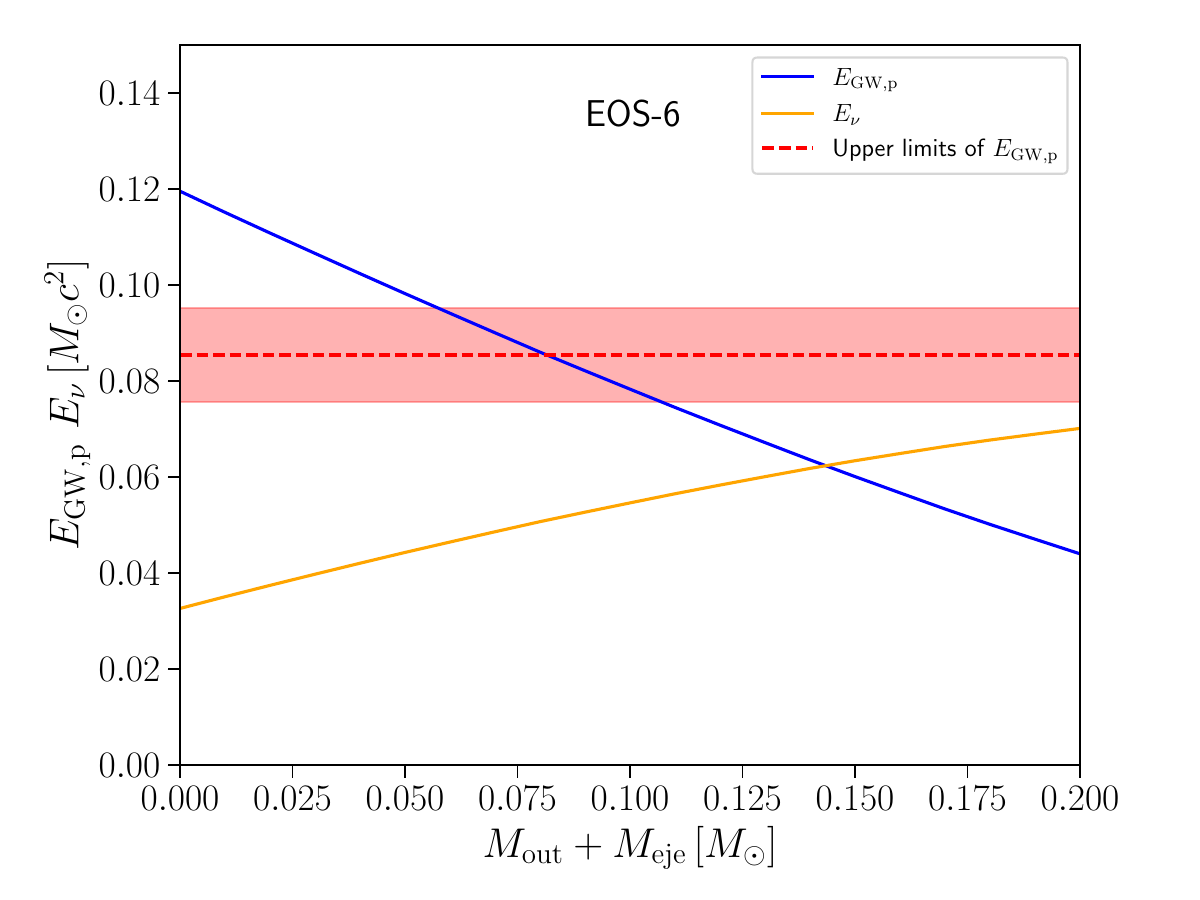}
        \\
        \includegraphics[width=0.48\textwidth]{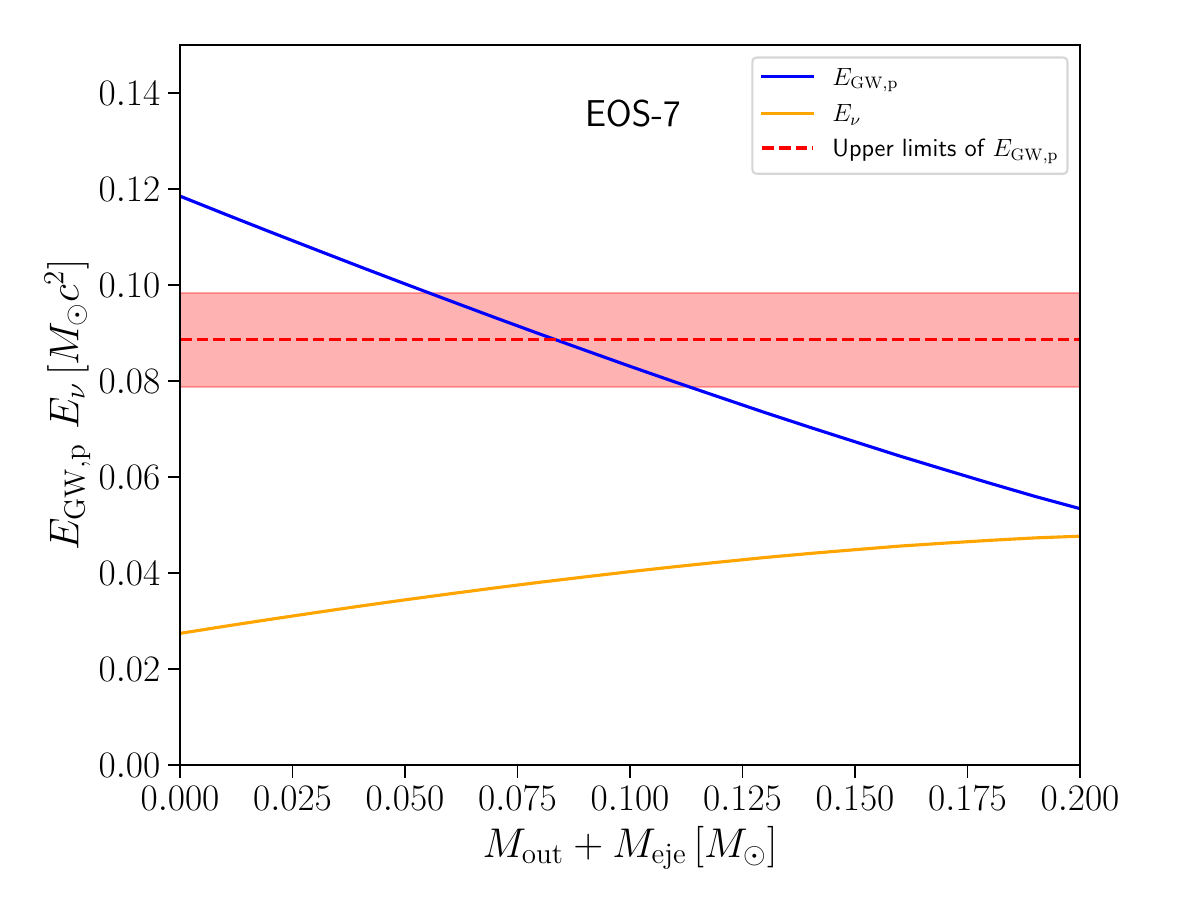}
        \quad
        \includegraphics[width=0.48\textwidth]{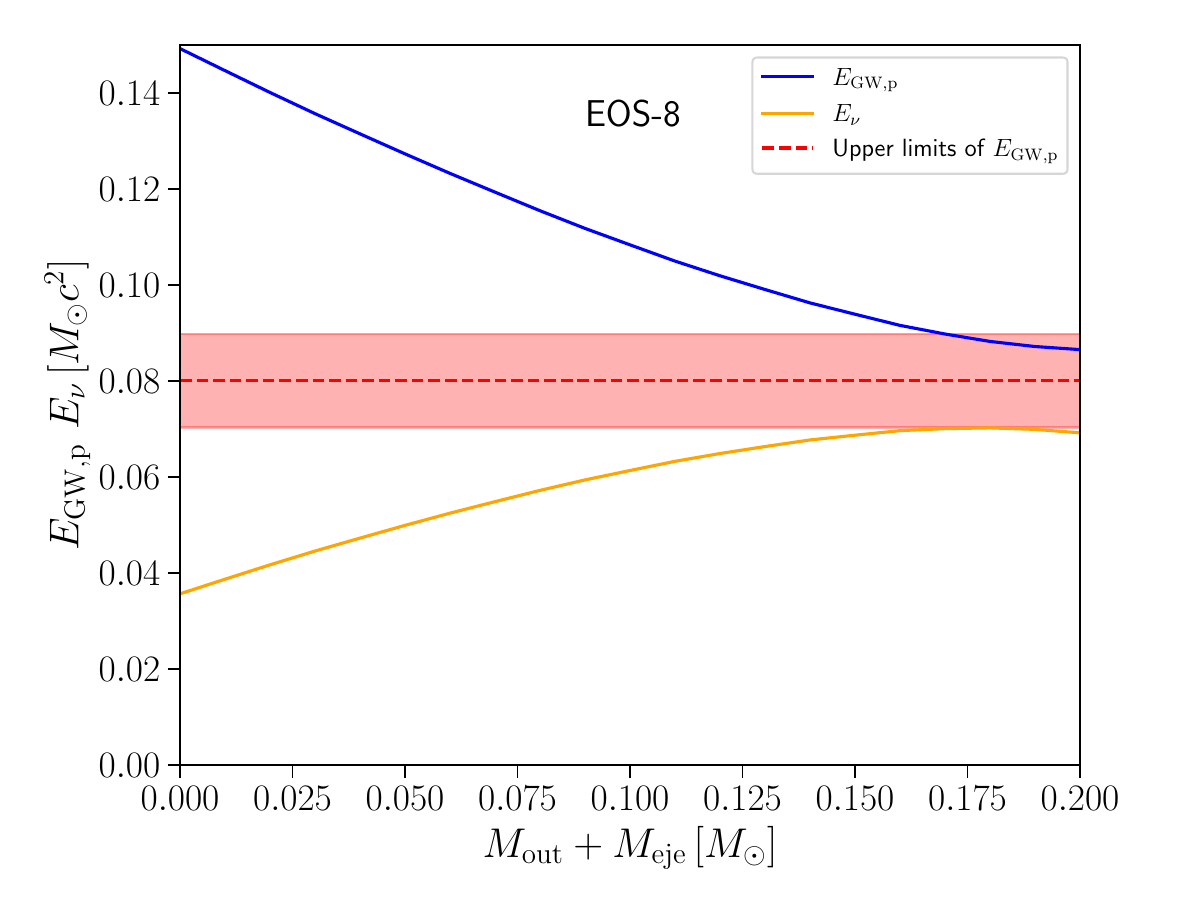}
        \\
        \includegraphics[width=0.48\textwidth]{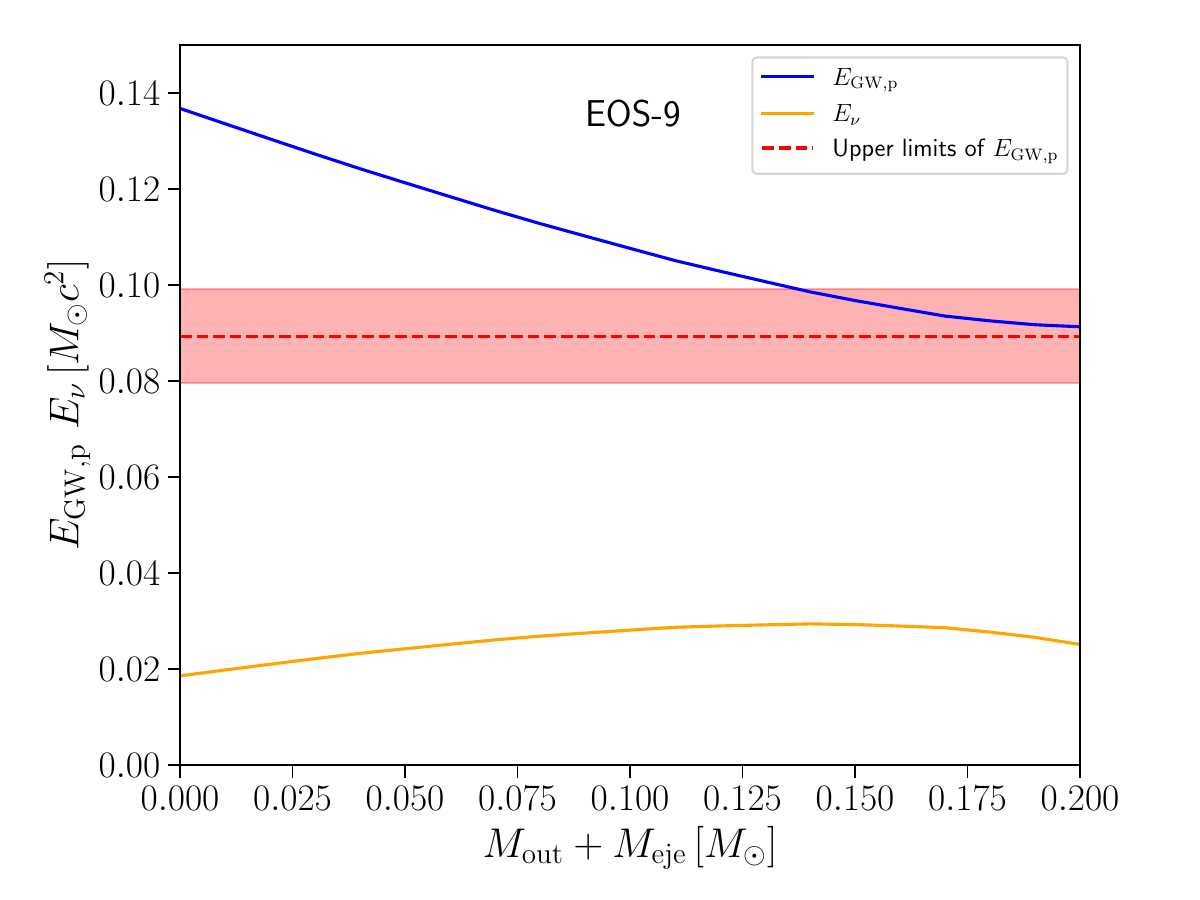}
        \quad
        \includegraphics[width=0.48\textwidth]{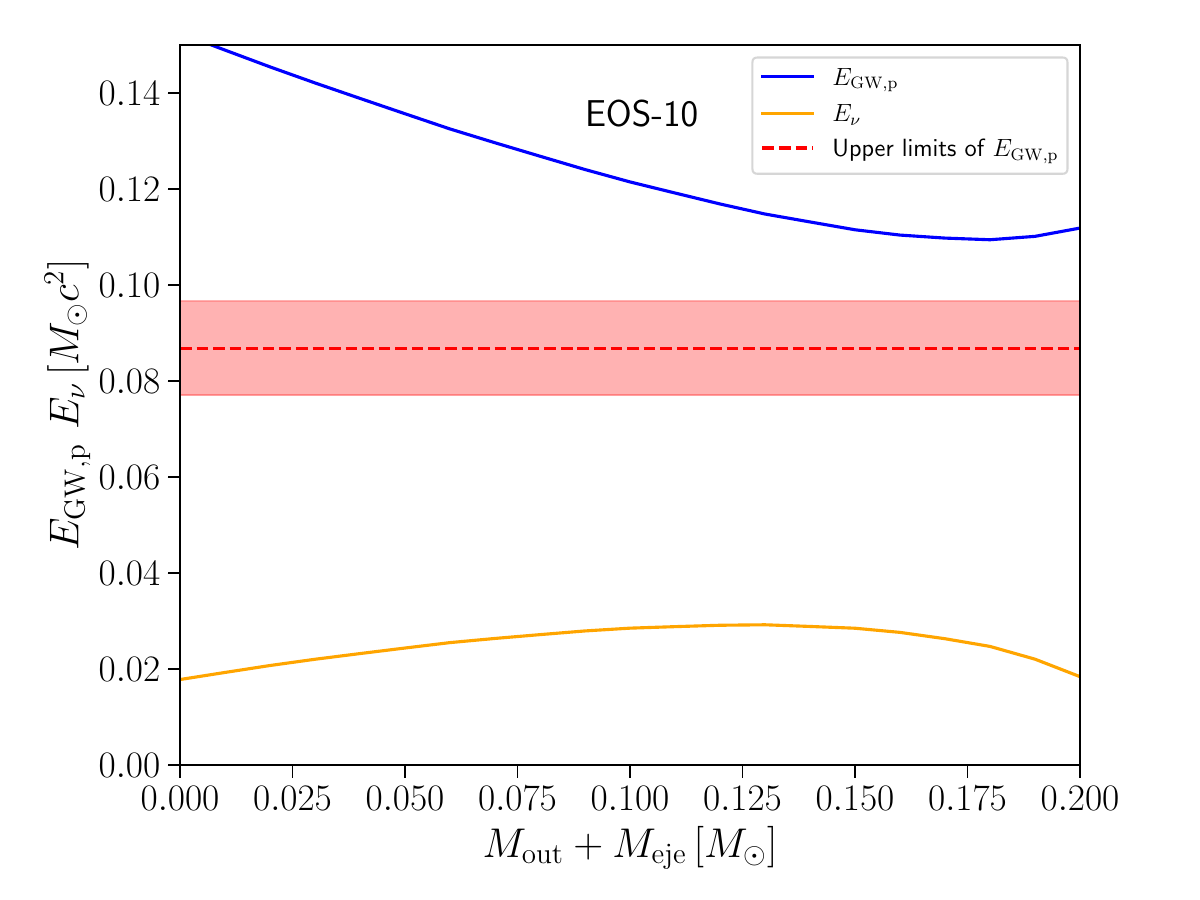}
	\end{center}
	\caption{The relation between $E_{\mathrm{GW,p}}$, $E_{\nu}$ and $M_{\mathrm{out}}+M_{\mathrm{eje}}$ for BQS of mass $1.37\,M_{\odot}$ and $1.37\,M_{\odot}$ with various QS EOSs. The blue and orange solid lines represent the energy radiated by gravitational waves and neutrinos, respectively. The horizontal dashed line represents the upper limits of the energy radiated by gravitational waves in the post-merger phase given by Eq. (\ref{19}) with the error bar indicated by the red shaded area. }
	\label{fig:results}
\end{figure*}

\subsection{Results with different $M_{\mathrm{TOV}}$}
\label{sec:same}
We first study the influence of $M_{\mathrm{TOV}}$ on the results. EOS-1 -- 7 have different $M_{\mathrm{TOV}}$ from $2.0\,M_{\odot}$ to $2.28\,M_{\odot}$, while the values of $\mathrm{sgn}(\lambda)\bar{\lambda}$ only vary from $-0.02$ to $1$.  Solutions that consistently satisfy the conservation laws of energy and angular momentum are shown in Table \ref{tab:results}, where the values of $M_{\mathrm{out}}+M_{\mathrm{eje}}$ are fixed to $0.05\,M_{\odot}$, $0.10\,M_{\odot}$, $0.15\,M_{\odot}$ and $0.20\,M_{\odot}$. The relation among $E_{\mathrm{GW,p}}$, $E_{\nu}$ and $M_{\mathrm{out}}+M_{\mathrm{eje}}$ is illustrated in Figure \ref{fig:results}. It can be observed from the Table \ref{tab:results} and the Figure \ref{fig:results} that an increase in $M_{\mathrm{out}}+M_{\mathrm{eje}}$ leads to a decrease in $E_{\mathrm{GW,p}}$. This is because the amount of angular momentum loss to trigger collapse is certain, and hence as $M_{\mathrm{out}}+M_{\mathrm{eje}}$ increases, more angular momentum is carried away by the torus and ejecta, and the value of $E_{\mathrm{GW,p}}$ decreases. In addition, as $M_{\mathrm{TOV}}$ increases, more angular momentum dissipation is needed to trigger the collapse. Therefore, the value of $E_{\mathrm{GW,p}}$ increases with the increase of $M_{\mathrm{TOV}}$ for the fixed $M_{\mathrm{out}}+M_{\mathrm{eje}}$.

It can also be seen in Table \ref{tab:results} and Figure \ref{fig:results} that for EOS-1, 2, 3, 4, 5, for which the maximum mass ranges from $2.01$ to $2.20\, M_{\odot}$, solutions that consistently meet our constraint can be found, and thus we have no reason to exclude these EOSs. In addition, As shown in Table \ref{tab:results}, the value of $M_f/M_{\mathrm{TOV}}$ for these EOSs is in the range of $1.1\lesssim M_f/M_{\mathrm{TOV}} \lesssim 1.3$ and is smaller than 1.4, which is the typical value of $M_{\mathrm{max,urot}}/M_{\mathrm{TOV}}$. Therefore, for these EOSs, the angular momentum left in the remnant QS is not large enough to reach the mass-shedding limit, and a supramassive remnant of the BQS merger could also collapse to a BH in a short timescale (before its kinetic energy is dissipated via electromagnetic radiation) due to the angular momentum dissipation.

As $M_{\mathrm{TOV}}$ increases, the dissipation of angular momentum required to trigger the collapse increases. For EOS-6, 7,  for which $2.252\,M_{\odot}\lesssim M_{\mathrm{TOV}}\lesssim2.284M_\odot$, solutions can also be obtained. However, a significant angular momentum dissipation through gravitational radiation or torus and ejecta is required. To determine whether this efficient dissipation of angular momentum is possible or not, NR simulations must be performed and the details of such NR simulations will be presented in another paper~\citep{Zhou2024}. 

\subsection{Results with the same $M_{\mathrm{TOV}}$ and the upper limit of $M_{\mathrm{TOV}}$}

The value of $\mathrm{sgn}(\lambda)\bar{\lambda}$ also influences the relation among $E_{\mathrm{GW,p}}$, $E_{\nu}$ and $M_{\mathrm{out}}+M_{\mathrm{eje}}$ while keeping $M_{\mathrm{TOV}}$ constant. For EOS-7 and EOS-8, ($M_{\mathrm{TOV}}\approx2.284\,M_{\odot}$), the great difference in $\mathrm{sgn}(\lambda)\bar{\lambda}$ leads to significantly different results. For EOS-7, with smaller $\mathrm{sgn}(\lambda)\bar{\lambda}$, we can easily find the self-consistent solutions, while for EOS-8 solutions can only be marginally found. The same trend is also found in the comparison of EOS-9 and EOS-10, ($M_{\mathrm{TOV}}\approx2.35\,M_{\odot}$). However, the difference in $\mathrm{sgn}(\lambda)\bar{\lambda}$ of EOS-9, 10 is not as significant as EOS-7,8, leading to a minor difference in the results. It can be seen that for EOS-9, the self-consistent solutions can be found within the uncertainty range, while for EOS-10 solutions can not be found because an unfeasibly high level of angular momentum dissipation is required to trigger the collapse.

To find the upper limit of $M_{\mathrm{TOV}}$ for the whole parameter space of QS EOSs, we need to calculate the results for EOSs with the same $M_{\mathrm{TOV}}$ but different $\mathrm{sgn}(\lambda)\bar{\lambda}$. For a given $M_{\mathrm{TOV}}$, if self-consistent solutions can not be found for any value of $\mathrm{sgn}(\lambda)\bar{\lambda}$, and then EOSs with this $M_{\mathrm{TOV}}$ can be excluded. In addition, EOSs with larger $M_{\mathrm{TOV}}$ can also be excluded because, as discussed before, EOS with larger $M_{\mathrm{TOV}}$ is more difficult to satisfy the constraint given by Eq. (\ref{19}). 

From the results of EOS-7--10, it can be seen that given a $M_{\mathrm{TOV}}$, EOSs with smaller $\mathrm{sgn}(\lambda)\bar{\lambda}$ are easier to find results satisfying the constraint. This is because as shown in Figure \ref{fig:f0}, for fixed $M_{\mathrm{TOV}}$ the decrease of $\mathrm{sgn}(\lambda)\bar{\lambda}$ leads to the increase of $f_0/f_{\mathrm{MS}}$, which leads to the increase of $M_{\mathrm{out}}+M_{\mathrm{eje}}$ for fixed $M_f$ and $M$. Therefore, for the same energy dissipation, the proportion of the dissipation through the torus and the ejecta is larger for EOS-7 and EOS-9, while for EOS-8 and EOS-10 more energy should be dissipated through gravitational waves and neutrinos, making $E_{\mathrm{GW}}$ exceed the upper limits given by Eq. (\ref{19}). Therefore, to satisfy the constraint given by Eq. (\ref{19}), an upper limit of $\mathrm{sgn}(\lambda)\bar{\lambda}$ can be obtained for a given $M_{\mathrm{TOV}}$.

\begin{figure}
	\begin{center}
		\includegraphics[height=70mm]{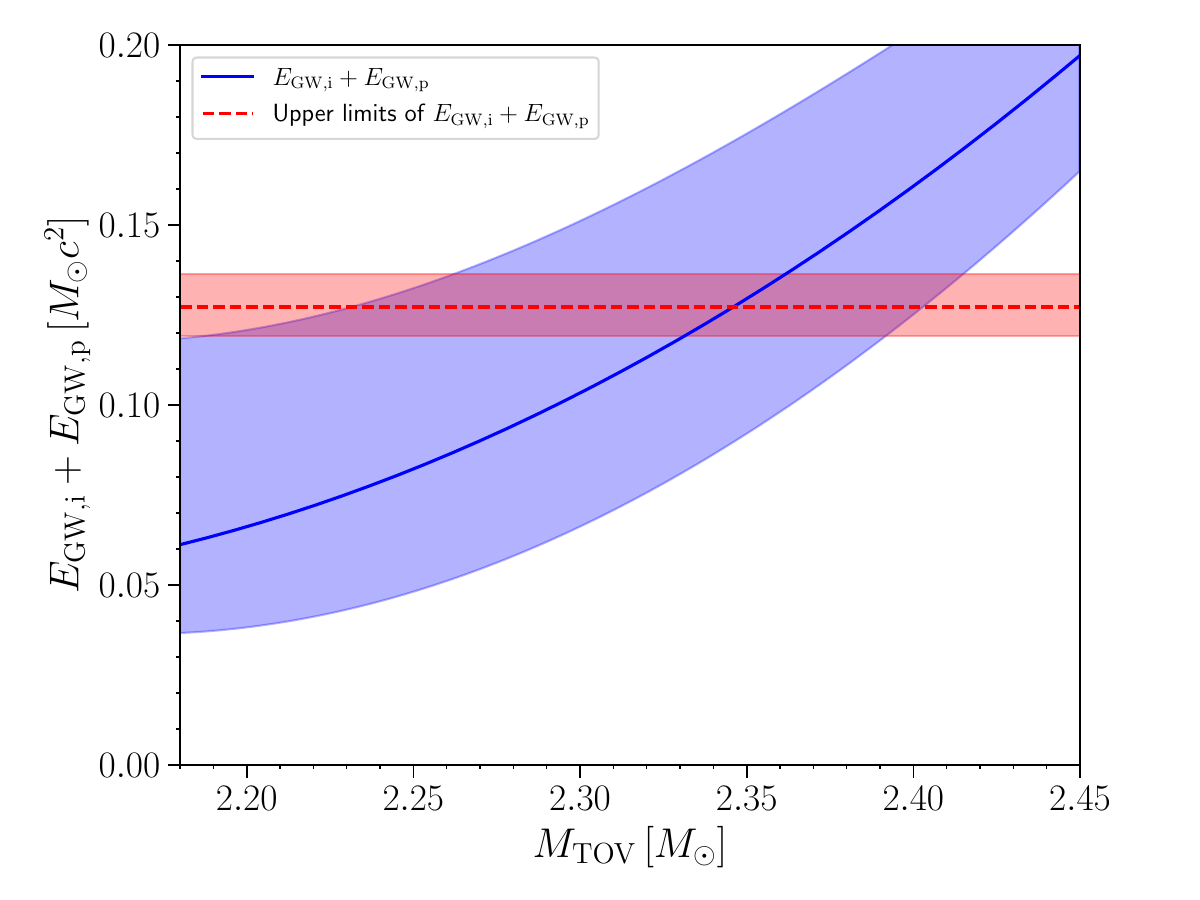}
   
	\end{center}
	\caption{Deriving the upper limit of $M_{\mathrm{TOV}}$ of QSs that can pass our consistency checks and satisfy $\Lambda_{1.4}\lesssim 800$ at the same time. The blue solid line represents the relation between $M_{\mathrm{TOV}}$ and the minimum $E_{\mathrm{GW,i}}+E_{\mathrm{GW,p}}$ to trigger the collapse derived by our semi-analytical method with the assumption that the total binary ADM mass $M=2.74\,M_{\odot}$, and the radius of the torus $R_{\mathrm{out}}=140\,$km for various EOSs with $\Lambda_{1.4}=800$ (We vary the value of $M_{\mathrm{out}}+M_{\mathrm{eje}}$ from $0$ to $0.2\,M_{\odot}$ to find the minimum value of $E_{\mathrm{GW}}$, which serve as the minimum $E_{\mathrm{GW,i}}+E_{\mathrm{GW,p}}$ to trigger the collapse.). The blue shaded area indicates the error arising from the uncertainties of $M$ ($2.74^{+0.04}_{-0.01}\,M_{\odot}$) and $R_{\mathrm{out}}$ ($140^{+60}_{-90}\,$km). The horizontal dashed line shows the upper limits of $E_{\mathrm{GW}}$ given by Eq.~(\ref{19}) with the assumption of $M=2.74\,M_{\odot}$. The red shaded area provides the error resulting from the uncertainties of $M$ and Eq. (\ref{19}). The intersection of these two lines determines the upper limit of $M_{\mathrm{TOV}}$ of QSs that can pass our consistency checks and satisfy $\Lambda_{1.4}\lesssim 800$ at the same time. }
	\label{fig:upper}
\end{figure}

It is suggested in~\citep{Zhang2021} that the most tight lower limits of $\mathrm{sgn}(\lambda)\bar{\lambda}$ are given by the constraint $\Lambda_{1.4}\lesssim 800$ from GW170817. Therefore, for a given $M_{\mathrm{TOV}}$, the upper limit and the lower limit of $\mathrm{sgn}(\lambda)\bar{\lambda}$ can be obtained by Eq.~(\ref{19}) and $\Lambda_{1.4}\lesssim 800$ respectively, and thus a definite upper limit of $M_{\mathrm{TOV}}$ can be derived by looking for the critical $M_{\mathrm{TOV}}$ with which the upper limit of $\mathrm{sgn}(\lambda)\bar{\lambda}$ given by the scheme demonstrated here is equal to the lower limit of it given by $\Lambda_{1.4}\lesssim 800$. It is equivalent to finding an EOS with which $\Lambda_{1.4}=800$  and Eq.~(\ref{19}) can be marginally satisfied.

Thus we scan over different $M_{\mathrm{TOV}}$ but all with $\Lambda_{1.4}=800$ to yield a definite answer. We vary the value of $M_{\mathrm{out}}+M_{\mathrm{eje}}$ from $0$ to $0.2\,M_{\odot}$ to find the minimum value of $E_{\mathrm{GW,i}}+E_{\mathrm{GW,p}}$ to trigger the collapse and compare it to Eq. (\ref{19}) to determine the upper limit of $M_{\mathrm{TOV}}$. In addition, the main uncertainties in our analysis are also considered. We vary the total binary ADM mass and the radius of the torus in their ranges of uncertainty and calculate the results ($M=2.74^{+0.04}_{-0.01}\,M_{\odot}$ with the $90\%$ credible level \citep{Abbott2017}, $R_{\mathrm{out}}$ varies from $50\,$km to $200\,$km due to the long-term viscous evolution as well as the angular momentum transport process~\citep{Fujibayashi2018,Fujibayashi2020}). Figure \ref{fig:upper} shows the main results: the blue solid line represents the relation between $M_{\mathrm{TOV}}$ and the minimum $E_{\mathrm{GW,i}}+E_{\mathrm{GW,p}}$ to trigger the collapse with the assumption that $M=2.74\,M_{\odot}$ and $R_{\mathrm{out}}=140\,$km, the horizontal red dashed line shows the upper limits of $E_{\mathrm{GW,i}}+E_{\mathrm{GW,p}}$ given by Eq.~(\ref{19}) with the assumption that $M=2.74\,M_{\odot}$ and the shaded areas indicate the uncertainties in $E_{\mathrm{GW,i}}+E_{\mathrm{GW,p}}$ associate with the uncertainties of $M$ and $R_{\mathrm{out}}$. The intersection of these two lines determines the critical $M_{\mathrm{TOV}}$ ($\approx2.35^{+0.07}_{-0.17}\,M_{\odot}$), and EOS with this critical $M_{\mathrm{TOV}}$ and $\Lambda_{1.4}=800$ can marginally satisfy the constraint given by Eq.~(\ref{19}). Therefore, it is illustrated that EOSs with $M_{\mathrm{TOV}}\gtrsim2.35^{+0.07}_{-0.17}\,M_{\odot}$ can not pass our consistency checks and satisfy the constraint $\Lambda_{1.4}\lesssim800$ at the same time. Note that in our analysis, we only vary $M_{\mathrm{out}}+M_{\mathrm{eje}}$ from 0 to $0.2\,M_{\odot}$, but $0.2\,M_{\odot}$ is not a solid upper limit of $M_{\mathrm{out}}+M_{\mathrm{eje}}$, and it is feasible especially for the unequal-mass case~\citep{Kiuchi2019,Kiuchi2024}. However, for EOSs with $M_{\mathrm{TOV}}>2.35^{+0.07}_{-0.17}\,M_{\odot}$, if $M_{\mathrm{out}}+M_{\mathrm{eje}}>0.2\,M_{\odot}$, such large mass loss would result in a remnant less massive than its $M_{\mathrm{TOV}}$ and becomes a stable remnant (see the lowest row in Table~\ref{tab:results}, when $M_{\mathrm{out}}+M_{\mathrm{eje}}=0.2\,M_{\odot}$, $M_{f}/M_{\mathrm{TOV}}\approx1.008$ is very close to 1, and thus for $M_{\mathrm{out}}+M_{\mathrm{eje}}>0.2\,M_{\odot}$ a stable remnant would be formed). Therefore, $M_{\mathrm{TOV}}$ has to be smaller than $2.35^{+0.07}_{-0.17}\,M_{\odot}$, as the very long-lived ($\sim 100\, \mathrm{s}$)/stable remnant can be excluded from the observation results of the GW170817 event~\citep{Margalit2017,Shibata2019}.

In conclusion, for the EOSs with $M_{\mathrm{TOV}}\lesssim 2.35^{+0.07}_{-0.17}\,M_{\odot}$, it is possible to pass our consistency checks and satisfy the constraint $\Lambda_{1.4}\lesssim800$ simultaneously, which is similar to the results for the BNS scenario in \citep{Shibata2019}. Although the value of $M_{\mathrm{max,urot}}/M_{\mathrm{TOV}}$ is larger for QSs, the angular momentum left in the remnant QS is always not large enough to reach the mass-shedding limit, and thus even the mass of remnant QS is smaller than $M_{\mathrm{max,urot}}$, the collapse can still take place before its kinetic energy is dissipated via electromagnetic radiation ($\sim 100\,\mathrm{s}$). On the other hand, despite the fact that interquark effects could significantly increase the $M_{\mathrm{TOV}}$ of QSs, QS EOSs with large $M_{\mathrm{TOV}}$ are disfavored by the electromagnetic counterparts of GW170817 event. It should be noted that the accurate evolution pictures of post-merger and the exact values of $E_{\mathrm{GW,p}}$, $E_{\nu}$ and $M_{\mathrm{out}}+M_{\mathrm{eje}}$ and the collapse behavior of the merger remnant can only be determined by NR simulations (e.g.,  \citep{Zhou2022,Zhou2024}), and our analysis can only offer relatively conservative constraint.

\subsection{Constraints on the  parameter space of QS EOSs}

\begin{figure}
	\begin{center}
		\includegraphics[height=70mm]{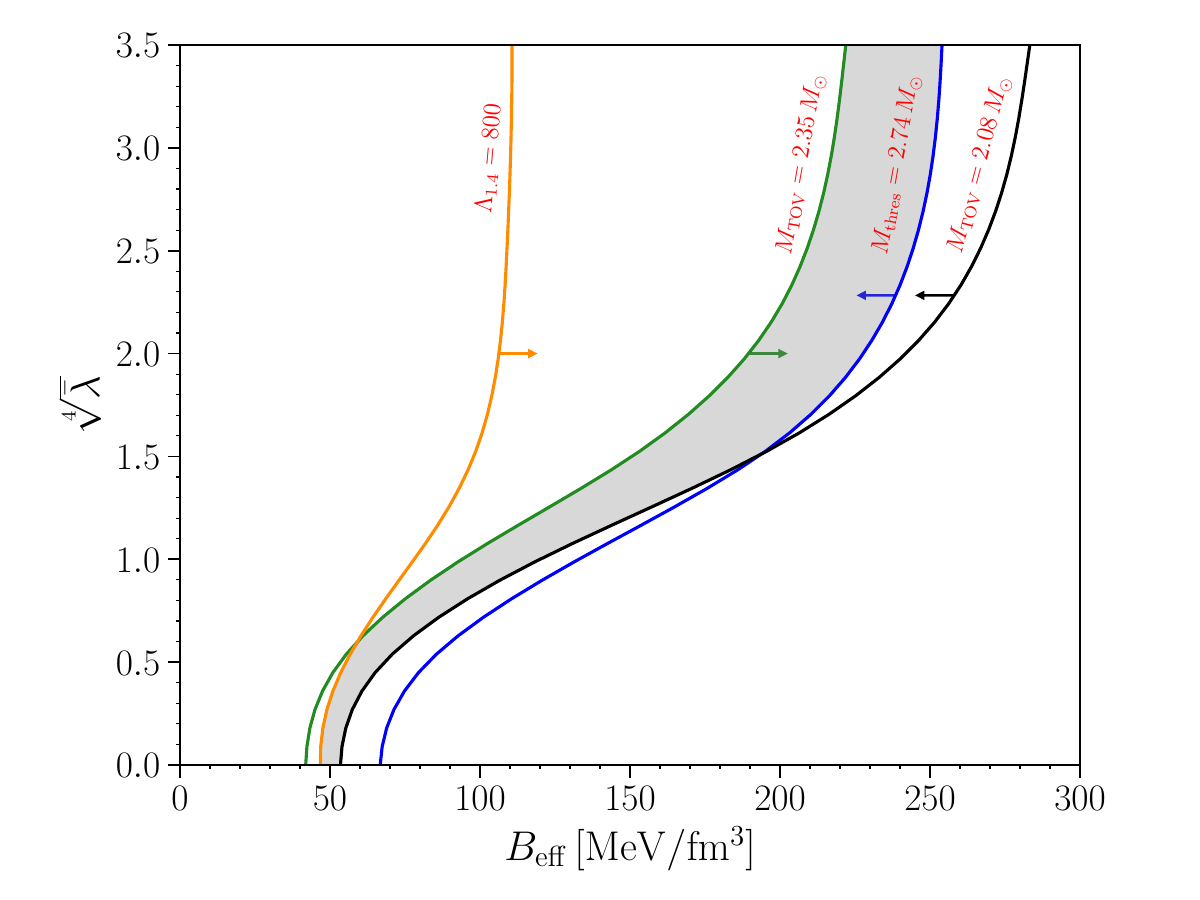}
   
	\end{center}
	\caption{Astrophysical constraints on the parameter space
in the ($B_{\mathrm{eff}},\bar{\lambda}^{1/4}$) plane (only positive $\lambda$ space is considered in this figure). The $\Lambda_{1.4}=800$, $M_{\mathrm{TOV}}=2.08,\ 2.35 \, M_{\odot}$ and $M_{\mathrm{thres}}=2.74\,M_{\odot}$ lines are plotted with different colors, which correspond to the GW170817 constraint ($\Lambda_{1.4}\lesssim800$~\citep{Abbott2017,Soumi2018}), the maximum pulsar mass constraint from PSR J0740+6620 ($M_{\mathrm{TOV}}\gtrsim 2.08\,M_{\odot}$~\citep{Fonseca2021, Cromartie2020}) and two constraints established by our semi-analytical method from the electromagnetic counterparts of GW170817 ($M_{\mathrm{TOV}}\lesssim2.35\,M_{\odot}$ and $M_{\mathrm{thres}} \gtrsim 2.74\,M_{\odot}$). The regions to the right of the orange and green solid lines satisfy the constraints $\Lambda_{1.4}\lesssim800$ and $M_{\mathrm{TOV}}\lesssim2.35\,M_{\odot}$, respectively. The regions to the left of the blue and black solid lines satisfy the constraints $M_{\mathrm{thres}}\gtrsim2.74\,M_{\odot}$ and $M_{\mathrm{TOV}}\gtrsim2.08\,M_{\odot}$, respectively.
The grey-shaded region indicates the allowed parameter space. 
}
	\label{fig:parameter}
\end{figure}
In this paper, constraints on $M_{\mathrm{TOV}}$ ($\lesssim2.35\,M_{\odot}$) and $M_{\mathrm{thres}}$ ($\gtrsim2.74\,M_{\odot}$) are established from the assumption that a delayed
collapse occurred before the kinetic energy of the merger remnant is dissipated via electromagnetic radiation
for the GW170817 event. Combining them with the current astrophysical constraints from GW170817 ($\Lambda_{1.4}\lesssim800$~\citep{Abbott2017,Soumi2018}) and PSR J0740+6620 ($M_{\mathrm{TOV}}\gtrsim 2.08\,M_{\odot}$~\citep{Fonseca2021, Cromartie2020}), we can well constrain the parameter space of QS EOSs and the resultant constrained parameter space is shown in Figure \ref{fig:parameter}, where the solid lines with different colors represent the different constraints and the blue line, $M_{\mathrm{thres}}=2.74\,M_{\odot}$ line, is calculated by Eq. (\ref{eq:UR}), which is solidified by our semi-analytical calculations
of equilibrium models and can be corroborated and refined by future NR simulations like~\citep{Zhou2021}. The GW190425~\citep{Abbott2020} event also makes constraint on the parameter of QS EOSs, but the constraint from it is looser compared to GW170817 constraint (see Figure 3 in \citep{Zhang2021}), and thus it is not displayed in Figure~\ref{fig:parameter}. 

As shown in Figure~\ref{fig:parameter}, for $\bar{\lambda}^{1/4}\gtrsim 0.6$ ($\bar{\lambda}^{1/4}\gtrsim 1.5$),  the lower (upper) bounds of $B_{\mathrm{eff}}$ are determined by $M_{\mathrm{TOV}}\lesssim2.35\,M_{\odot}$ ($M_{\mathrm{thres}}\gtrsim2.74\,M_{\odot}$). On the other hand, for $\bar{\lambda}^{1/4}\lesssim 0.6$ ($\bar{\lambda}^{1/4}\lesssim 1.5$),  the lower (upper) bounds of $B_{\mathrm{eff}}$ are determined by $\Lambda_{1.4}\lesssim800$ ($M_{\mathrm{TOV}}\gtrsim2.08\,M_{\odot}$). It should be noted that only positive $\lambda$ space is considered in Figure~\ref{fig:parameter}. The parameter space of QS EOSs with negative $\lambda$ has been well-constrained by $\Lambda_{1.4}\lesssim800$ and $M_{\mathrm{TOV}}\gtrsim2.08\,M_{\odot}$~\citep{Zhou2018} and the investigation in this paper do not impose a further constraint on it.
Therefore, the investigation of the post-merger evolution for the GW170817 event can make tight constraints on QS EOSs with the large $\mathrm{sgn}(\lambda)\bar{\lambda}$.

\section{CONCLUSIONS AND DISCUSSIONS}
\label{sec:conclusion}

We semi-analytically investigate the post-merger evolution of the BQS merger. With the help of the EOB method and considering the angular momentum dissipation process, we obtain the angular momentum left in the merger remnant at the time of the merger. The amounts of three main dissipation mechanisms in the post-merger phase are determined by the conservation laws of baryon number, energy and angular momentum. In addition, we determine the stability of the remnant QS by constructing the equilibrium models of rotating QSs. Through this semi-analytical method, we investigate both prompt collapse and delayed collapse cases and set an upper bound on QSs' $M_{\mathrm{TOV}}$ through the GW170817 event. 

Assuming GW170817 originated from a BQS merger, we analyze the post-merger evolution for GW170817 in the BQS scenario. Our analysis suggests that a QS with $M_{\mathrm{TOV}}>2.0\,M_{\odot}$, leading to $M_{\mathrm{max,urot}}>2.8\,M_{\odot}$, can collapse to a BH before its kinetic energy and angular momentum are dissipated via electromagnetic radiation ($\sim 100\,\mathrm{s}$) because due to the angular momentum dissipation of mass outflows, neutrinos and gravitational waves, the angular momentum left in the remnant QS might not be large enough to sustain the additional self-gravity of the supramassive QS. Moreover, we also systematically investigate the influence of $M_{\mathrm{TOV}}$ and $\mathrm{sgn}(\lambda)\bar{\lambda}$ on the results of our consistency checks. Finally, an upper limit of $M_{\mathrm{TOV}}$ is obtained, and only QS EOSs with $M_{\mathrm{TOV}}\lesssim2.35^{+0.07}_{-0.17}\,M_{\odot}$ can pass our consistency checks and satisfy $\Lambda_{1.4}\lesssim800$ simultaneously. Therefore, although considering interquark effects can make $M_{\mathrm{TOV}}$ of QSs very large, these QS EOSs with large $M_{\mathrm{TOV}}$ are unlikely to pass the constraint of the electromagnetic counterparts of the GW170817 event. It should be noted that our derived constraints in this paper are valid for the EOSs based on the IQM model \citep{Zhang2021} in the one-family scenario, while other cases of different quark models (such as quark-mass density-dependent model ~\citep{fowler1981confinement,chakrabarty1989strange,Peng:1999gh,Xia:2014zaa} and NJL model~\citep{Buballa:2003qv,Yuan:2022dxb,Yuan:2023dxl,Xia:2024wpz}) or different scenarios (such as the two-family scenario ~\citep{Berezhiani2003,Bombaci2004,Drago2004})  are not considered, but analyses could be done in a similar way.

Although there are many unknowns in the electromagnetic counterparts associated with BQS mergers, our constraint $M_{\mathrm{TOV}}\lesssim2.35\,M_{\odot}$ is only based on the exclusion of the very long-lived ($\sim 100\,\mathrm{s}$)/stable remnant that injects a large amount of its kinetic energy into the outflows from the observation results of the GW170817 event~\citep{Margalit2017, Shibata2017,Shibata2019}. In comparison, the assumption that the short $\gamma$-ray burst is powered by the spinning BH-torus system~\citep{Ruiz2018} may be inappropriate considering the simulations in~\citep{Kiuchi2024,Bamber2024}, where it is indicated that SMNS and HMNS may be able to serve as the central engine of the short $\gamma$-ray burst. Additionally, as discussed in~\citep{Shibata2019} and the current paper, the assumption that the remnant NS/QS is at the mass-shedding limit is also improper, and thus for both NSs and QSs, $M_{\mathrm{TOV}}$ can only be weakly constrained as $M_{\mathrm{TOV}}\lesssim2.35\,M_{\odot}$. Moreover, our analysis can also connect to the multi-messenger observations. In this paper, we only employ the constraint of the energy of gravitational waves obtained by NR simulations. In the future, with observations of post-merger gravitational waves and neutrinos, the values of $E_{\mathrm{GW,p}}$ and $E_{\mathrm{\nu}}$ can be further constrained by observations. As a result, the constraints for EOSs of compact stars can be better imposed and the central engine of the short $\gamma$-ray burst can be understood more deeply.

\acknowledgements

We thank the members of the Relativistic AstroParticle Physics group at Huazhong
University of Science and Technology and the Computational Relativistic
Astrophysics division at Max Planck Institute for Gravitational Physics
(Potsdam) for their helpful discussions. E. Z. acknowledges the support of the
National SKA Program of China No. 2020SKA0120300 and NSFC Grant No. 12203017.
C.~Z. is supported by the Jockey Club Institute for Advanced Study at The Hong
Kong University of Science and Technology. J.~Z. is supported by the NSFC Grant
No.~12147177 and the
``LiYun'' Postdoctoral Fellowship of Beijing Normal University. 
This work was in part supported by the Grant-in-Aid for Scientific Research (grant Nos. 23H01172, 23K25869) of Japan MEXT/JSPS.

\bibliographystyle{apsrev4-1}
\bibliography{ref}

\end{document}

%% file: kigou.tex
\newcommand{\beq}{\begin{equation}} 
\newcommand{\eeq}{\end{equation}} 
\newcommand{\beqn}{\begin{eqnarray}} 
\newcommand{\eeqn}{\end{eqnarray}}

\newcommand{\zD}{{\raise1.0ex\hbox{${}^{\ \circ}$}}\!\!\!\!\!D}
\newcommand{\alone}{{\raise0.5ex\hbox{${}^{\ 1}$}}\!\!\!\!\alpha}

\newcommand{\nalam}{\mathrel{\raise0.9ex\hbox{$^\lambda$}\mkern-14mu
\lower0.0ex\hbox{$\nabla$}}}

\newcommand{\zeroD}{{\raise1.0ex\hbox{${}^{\ \circ}$}}\!\!\!\!\!D}
\newcommand{\zLap}{{\raise1.0ex\hbox{${}^{\ \circ}$}}\!\!\!\!\Delta}
\newcommand{\zna}{{\raise1.0ex\hbox{${}^{\ \circ}$}}\!\!\!\!\!\nabla}
\newcommand{\zS}{{\raise1.0ex\hbox{${}^{\ \circ}$}}\!\!\!\!\!S}

